\documentclass[journal]{IEEEtran}
\usepackage{amsthm}
\usepackage{empheq}
\usepackage{cancel}
\usepackage{mdframed,xcolor}
\usepackage{tcolorbox}
\usepackage{placeins}
\usepackage{relsize}
\usepackage{setspace}
\usepackage{stmaryrd}
\usepackage{xcolor}
\usepackage{url}
\usepackage{multirow}
\usepackage{amssymb}
\usepackage{amsmath}
\usepackage[end]{algpseudocode}
\usepackage{algorithm}
\usepackage{amsfonts}
\usepackage{diagbox}
\usepackage{booktabs}
\usepackage{filecontents}
\usepackage{mathtools}
\usepackage{caption}
\usepackage{subcaption}
\usepackage{enumitem}
\usepackage{xfrac}
\usepackage{nicefrac}
\usepackage{soul}
\usepackage{algorithm}
\algdef{SE}{Begin}{End}{\textbf{begin}}{\textbf{end}}
\setlist[itemize]{leftmargin=*}
\usepackage{xcolor}
\definecolor{rv1}{rgb}{1.0, 0.44, 0.37}
\definecolor{rv2}{rgb}{0.4, 1.0, 0.0}
\definecolor{rv3}{rgb}{0.0, 0.75, 1.0}
\definecolor{rvt}{rgb}{0.75, 0.75, 0.75}
\definecolor{my-blue}{cmyk}{0.1, 0.0, 0.0, 0.0, 1.00}
\usepackage{soul}

\newtheoremstyle{exampstyle}
{7pt} 
{7pt} 
{\itshape} 
{} 
{\bfseries} 
{.} 
{.5em} 
{} 
\theoremstyle{exampstyle}

\newcommand{\algrule}[1][.2pt]{\par\vskip.5\baselineskip\hrule height #1\par\vskip.5\baselineskip}

\newcommand{\ab}{{\bf a}}

\newcommand{\phib}{{\mbox{\boldmath $\phi$}}}

\newcommand{\mub}{{\mbox{\boldmath $\mu$}}}
\newcommand{\thetab}{{\mbox{\boldmath $\theta$}}}

\newcommand{\alphab}{{\mbox{\boldmath $\alpha$}}}

\newcommand{\vb}{{\mathbf v}}

\newsavebox\mybox

\begin{document}

\title{Meta Reinforcement Learning for Resource Allocation in Aerial Active-RIS-assisted Networks with Rate-Splitting Multiple Access}

\author{Sajad Faramarzi, Sepideh Javadi, Farshad Zeinali, Hosein Zarini, Mohammad Robat Mili, \\Mehdi Bennis, \textit{Fellow, IEEE}, Yonghui Li, \textit{Fellow, IEEE}, and Kai-Kit Wong, \textit{Fellow, IEEE}
\thanks{S. Faramarzi (e-mail: sajad.faramarzi$1397$@gmail.com) is with the School of Electrical Engineering, Iran University of Science $\&$ Technology, Tehran, Iran. S. Javadi (e-mail: sepideh.javadi@piais.ir), F. Zeinali (e-mail: farshad.zeinali@piais.ir), and M. Robat Mili (e-mail: mohammad.robatmili@gmail.com) are with the Pasargad Institute for Advanced Innovative Solutions (PIAIS), Tehran, Iran. H. Zarini (e-mail: hosein.zarini68@sharif.edu) is with the Dept. of Computer Engineering, Sharif University of Technology, Tehran, Iran. Mehdi Bennis (e-mail:mehdi.bennis@oulu.fi) is with the Department of electrical Engineering, University of Oulu, Oulu, Finland. Yonghui Li (e-mail: yonghui.li@sydney.edu.au) is with the School of Electrical and Information Engineering, University of Sydney, Sydney, NSW 2006, Australia. K. K. Wong (e-mail: kai-kit.wong@ucl.ac.uk) is affiliated with the Department of Electronic and Electrical Engineering, University College London, Torrington Place, WC 1E 7JE, United Kingdom. He is also affiliated with Yonsei Frontier Lab, Yonsei University, Seoul, Korea.}}

\markboth{}%
{Shell \MakeLowercase{\textit{et al.}}: Bare Demo of IEEEtran.cls for IEEE Journals}

\maketitle
\begin{abstract}
Mounting a reconfigurable intelligent surface (RIS) on an unmanned aerial vehicle (UAV) holds promise for improving traditional terrestrial network performance. Unlike conventional methods deploying passive RIS on UAVs, this study delves into the efficacy of an aerial active RIS (AARIS). Specifically, the downlink transmission of an AARIS network is investigated, where the base station (BS) leverages rate-splitting multiple access (RSMA) for effective interference management and benefits from the support of an AARIS for jointly amplifying and reflecting the BS's transmit signals. Considering both the non-trivial energy consumption of the active RIS and the limited energy storage of the UAV, we propose an innovative element selection strategy for optimizing the on/off status of RIS elements, which adaptively and remarkably manages the system's power consumption. To this end, a resource management problem is formulated, aiming to maximize the system energy efficiency (EE) by jointly optimizing the transmit beamforming at the BS, the element activation, the phase shift and the amplification factor at the RIS, the RSMA common data rate at users, as well as the UAV's trajectory. Due to the dynamicity nature of UAV and user mobility, a deep reinforcement learning (DRL) algorithm is designed for resource allocation, utilizing meta-learning to adaptively handle fast time-varying system dynamics. Simulations indicate that incorporating an active RIS at the UAV leads to substantial EE gain, compared to passive RIS-aided UAV. 
We observe the superiority of the RSMA-based AARIS system in terms of EE, compared to existing approaches adopting non-orthogonal multiple access (NOMA).
\end{abstract}
\vspace*{-0.1em}
\vspace*{-0.1em}
\IEEEpeerreviewmaketitle
\vspace{-1.05em}
\section{Introduction}
\subsection{Background and Incentives}
The upcoming sixth-generation (6G) wireless networks are anticipated to support massive connectivity, due to the roll-out of a huge number of emerging Internet-of-Things (IoT) devices. This achievement, however, will be realized at the expense of an explosion in energy consumption and data rate of existing wireless networks~\cite{massive}. In light of this fact, the deployment of more advanced communication technologies is instrumental in enhancing the performance of existing networks, from both the spectral efficiency (SE) and energy efficiency (EE) perspectives~~\cite{near1}, \cite{near2}, \cite{near3}.
\par To achieve this goal, reconfigurable intelligent surfaces (RISs) as a revolutionary technology customize the wireless propagation environment by regulating their metasurface elements, reflecting incident signals, and harnessing inter-user interference~\cite{Hosein3}. For instance, in various practical mobile urban cellular environments, line-of-sight (LoS) links between the base station (BS) and users may not be available or can be obstructed by moving obstacles. Through adaptively bypassing these environmental obstacles and establishing virtual LoS links, RISs play a pivotal role in enhancing the network accessibility, thereby improving SE and EE in wireless networks \cite{multiplicative_fading}. In literature, RISs are mostly assumed to be deployed at fixed locations and typically operate in passive mode by only reflecting the incident signals without any amplification mechanism. However, this approach, i.e., the so-called multiplicative fading effect, inherent in the associated dual-hop channels \cite{Hosein3, multiplicative_fading, Hosein-TCOM} significantly weakens the signal strength in conventional RIS-enabled systems that rely on passive beamforming. This detrimental phenomenon results in substantial performance degradation in signal detection at the receiver \cite{es4}. In contrast, at the expense of increased power consumption, active RISs exert an amplification mechanism while reflecting incident signals. This approach effectively relieves the multiplicative fading effect, enabling signals to be well detected at the desired destinations~\cite{es7}. 
\par Recently, by mounting the RIS on an unmanned aerial vehicle (UAV), an aerial RIS (ARIS) system has been introduced and attracted extensive interests in the literature (see \cite{Surv1} and the references therein). This efficient system benefits from the combined advantages of both UAV and RIS in extending network coverage, as well as improving SE and EE, compared to the static terrestrial networks with stationary RISs. In deed, the success of ARIS systems highly relies on the corresponding multiple access scheme, as well as the resource allocation strategy, implemented. 
From the multiple access standpoint, rate-splitting multiple access (RSMA), as the most prominent scheme in wireless networks, has been introduced in~\cite{RSMA2}, which remarkably enhances both the SE and EE of non-orthogonal multiple access (NOMA), thanks to more efficient interference management. Surveying the literature in ARIS systems showcases various efforts investigating NOMA \cite{NOMA1,NOMA2}, which highlight performance enhancements over conventional orthogonal multiple access (OMA) schemes. 
For instance, the authors of \cite{NOMA1} proposed a resource allocation framework with NOMA aimed at minimizing the average power consumption of this system, whereas in \cite{NOMA2}, the downlink coverage of an ARIS system was analyzed and the corresponding coverage probabilities were derived. On the other hand, from the resource allocation perspective, two classes of methodologies have been investigated in literature to optimize ARIS systems, namely classical convex optimization-based ones~\cite{optimization1,optimization2}, and recent learning-driven ones~\cite{learning1,learning2}.
Despite presenting sub-optimal solutions, the former category relies on advanced mathematical transformations for approximating the non-convex resource allocation design problem by a convex counterpart, such that the latter can be solved by off-the-shelf solver tools, such as CVX~\cite{CVX}. In particular, this class of solutions is mainly characterized by significant computational complexity and time-consuming convergence. Comparatively, deep reinforcement learning (DRL) methods as model-free solutions frameworks (belonging to the latter category) have been widely applied in the literature for dynamic radio resource allocation in ARIS systems~\cite{Surv2}. Advantageous in jointly optimizing a multitude of system variables, this class of solutions does not rely on decomposition, which is widely leveraged in the prior category. Aside from this, DRL methods are mostly recognized as a real-time solution to dynamic resource allocation problems, e.g., \cite{learning1,learning2}. This advantage nonetheless, depends on the size of the action space in DRL networks, as well as the level of dynamicity of the wireless environment, within which they operate. 
\vspace*{-1em}
\subsection{Related Works}
\vspace{-0.2em}
The related works are classified into four categories, as follows:

1) \textbf{\ RIS-based systems:} In recent years, RIS-aided communications have been widely studied in different scenarios. More specifically, the authors in \cite{RelatedWork1} proposed an active RIS-assisted multi-user communication system, where the phase shift and amplification factor at the RIS, along with the transmit beamforming at the BS, were jointly optimized to minimize the system power consumption. 
Also, a hybrid active-passive RIS-aided system was studied in \cite{RelatedWork2}, by optimizing the number of active/passive elements to maximize the overall system EE.
Besides, in \cite{RelatedWork3}, the authors proposed a quadratic transform-based fractional programming technique
to investigate the overall EE of an active RIS-assisted multi-user system by jointly optimizing the transmit beamforming at the BS, as well as the amplification factor and phase shift at the RIS.
Furthermore, a desired trade-off between the overall system SE and EE was characterized in \cite{RelatedWork4}, assuming an active RIS-aided network adopting RSMA. In addition, the overall throughput maximization problem of an RIS-aided multi-carrier NOMA system was investigated in \cite{Javadi}, where the authors applied an alternating optimization (AO) method to acquire a sub-optimal solution to the non-convex resource allocation problem.

2) \textbf{\ UAV-based systems:} UAVs, as another pioneer communication
technology, have been significantly explored in the last
decade \cite{RelatedWork5}, \cite{RelatedWork6}. For instance, the authors in \cite{RelatedWork5} proposed a DRL-based technique to maximize the overall system EE of a UAV network by considering the communication coverage, fairness, energy consumption, and connectivity. Besides, relying on multi-armed bandit, the problem of maximizing overall system EE was modelled in \cite{RelatedWork6}, where the authors proposed an upper confidence bound-based algorithm via DRL. Recently, the authors of \cite{Hosein-ICC} adopted a DRL-based resource allocation methodology, so as to coordinate between UAVs in downlink transmission of a visible light communication (VLC) system. However, \cite{RelatedWork5,RelatedWork6,Hosein-ICC} consider the UAV as a flying BS to serve the network and don't utilize the terrestrial stations such as BS.

3) \textbf{\ ARIS-based systems:} These systems leverage UAV as a passive relay that reflects the incident signals of the BS towards users with aid of a mounted RIS. Maximizing the overall SE and EE of an ARIS system was studied in \cite{RelatedWork7}, by jointly optimizing the active beamforming at the BS, passive beamforming at the RIS, and the motion trajectory of the UAV.
In addition, the performance of an ARIS system in a smart railway network was investigated in \cite{RelatedWork8}, where maximizing the minimum SE of trains was targeted by jointly optimizing the trajectory at the UAV and the phase shift at the RIS. Besides, in \cite{RelatedWork9}, the application of deep
deterministic policy gradient (DDPG)-based DRL was analyzed to maximize the sum rate of a NOMA-enabled ARIS system, by jointly optimizing the power allocation at the BS, the phase shift at the RIS, and the horizontal position at the UAV.
Moreover, exploiting a deep neural network (DNN), the authors of \cite{RelatedWork10} proposed a resource allocation framework in an ARIS system with NOMA to minimize the average total system power consumption by jointly designing the three dimensional (3D) trajectory at the UAV, as well as the phase shift at the RIS.

4) \textbf{RSMA-based systems:} RSMA was previously shown in literature to be an energy-efficient multiple access scheme. For instance, aimed at maximizing the overall system EE, the performance of an active RIS-aided millimeter-wave (mmWave) hybrid antenna array system was assessed in \cite{RelatedWork12}, where a 2-layer RSMA was applied for effective user grouping.
Furthermore, in \cite{RelatedWork13}, the authors investigated the overall EE of the RSMA and NOMA in downlink transmission of a mmWave UAV network. It was numerically shown that the RSMA outperforms NOMA in terms of overall system EE. 
\vspace*{-1em}
\subsection{Research Challenges and Contributions}
In practice, to extend the coverage area and guarantee higher levels of quality-of-service (QoS) of the existing ARIS systems~\cite{Surv1,Surv2}, there is a need to elevate these systems architecturally. Regarding the advantages brought by active RISs over the traditional passive type~\cite{multiplicative_fading}, substituting the passive RIS at the UAV with an active one, turns out to be a promising approach to that purpose. However, due to a higher power consumption level at the active RISs, it is of utmost importance to devise an energy efficient resource allocation mechanism that accommodates the limited energy budget of the UAV. Additionally, our investigations  unveils that multiple access schemes in ARIS systems have never gone beyond NOMA~\cite{NOMA1,NOMA2}. Thanks to exploiting advanced interference cancellation techniques, utilization of RSMA~\cite{RSMA1, SLIPT_RSMA} is expected to further progress the performance of existing ARIS systems over its NOMA counterpart, provided that a meticulous resource allocation scheme is applied.
Unfortunately, existing resource allocation mechanisms adopted in ARIS systems, especially those based on DRL, exhibit shortcomings, when faced with substantial environmental dynamics, such as the mobility of the UAV and users. This challenge is attributed to the high dynamism in wireless environment, leading to frequent changes during the training and validation stages of DRL and thereupon a mismatch between these two phases~\cite{Meta1}. Consequently, prior DRL-based solutions, e.g.,~\cite{learning1,learning2}, are not applicable in highly dynamic scenarios. 

In this paper, we optimize the performance of current ARIS systems~\cite{Surv1,Surv2} by implementing an active RIS on the UAV, termed as aerial active RIS-assisted (AARIS) system. Regarding the considerable power consumption of an active RIS compared to its passive counterpart and also the limited battery storage of the UAV, we devise an element selection mechanism at the active RIS, which optimizes the on/off status of each element and adaptively controls its power consumption. Compared to~\cite{NOMA1,NOMA2}, our proposed AARIS system enables enhanced interference management by adopting RSMA, which enables improvement in overall system SE and EE. In addition, we design a meta reinforcement learning framework for resource allocation, with remarkable progress in generalization ability of DRL-based resource allocation schemes~\cite{learning1,learning2} by quickly adapting to the system dynamicity and recent environmental changes~\cite{Meta1}. Specifically, the main contributions of this paper can be outlined as follows:
\begin{itemize}
    \item This paper focuses on the performance of an AARIS system. In contrast to the ARIS literature (see \cite{Surv1,Surv2} and the references therein), where a passive RIS is mounted by the UAV, we consider an AARIS system for jointly amplifying and reflecting the downlink transmit signals of the BS toward users. We also improve the performance of NOMA-enabled ARIS systems \cite{NOMA1,NOMA2}, by adopting RSMA at the BS for more efficient interference management.
    \item Regarding the considerable energy consumption of the active RIS and also the battery limitation of the UAV, we introduce an element selection mechanism at the RIS to effectively control the power consumption of the AARIS system. We formulate the resource allocation design as an optimization problem, aiming at maximizing the overall system EE by optimizing the transmit beamforming at the BS, element selection, as well as the phase shift and amplification factor at the RIS, the RSMA common data rate at users, and the motion trajectory at the UAV in a joint manner.
    \item A DRL-based solution framework is proposed to address this resource allocation problem, which adopts a modified-soft actor-critic (modified-SAC) algorithm for optimizing the RIS element selection as a discrete optimization variable, as well as a twin delayed deep deterministic policy gradient (TD3) algorithm for optimizing other continuous-domain optimization variables~\cite{SAC_algorithm}, \cite{TD3_algorithm}. \textcolor{black}{Concerning the high dynamicity of this system (due to the mobility of the UAV and users), we fine-tune both DRL algorithms via meta-learning, enabling them to quickly adapt to recent system dynamic changes and new tasks.}
    \item Simulation results verify the effectiveness of the proposed AARIS system and the meta reinforcement learning resource allocation mechanism presented in this paper. In particular, the proposed AARIS system with an active RIS deployed on the UAV, achieves overall system EE gain over the ARIS system with a passive RIS deployed on the UAV \cite{Surv1}. Furthermore, thanks to efficient interference management enabled by RSMA, the proposed AARIS system in this paper outperforms the ARIS system \cite{NOMA1} which employs NOMA, in terms of overall system EE. Apart from the system model, our proposed meta reinforcement resource allocation scheme is superior to those conventional DRL-based schemes \cite{learning1,learning2,learning3}, by taking the advantage of further generalization ability and adaptness to the upcoming system danymics.  
\end{itemize}
\par \textit{Notations}: In this paper, the following notations are adopted. The capital boldface letters represent matrices, while the lower case of bold and normal letters denote vectors and scalars, respectively. An identity matrix is represented by $\mathbf{I}$. $\mathbb{C}^{x\times y}$ represents an $x$-by-$y$ complex-valued matrix and the superscript $H$ denotes the conjugate transpose of a matrix. 
The notations $\mathbb{E}\left[\cdot \right]  $ and
$\mathrm{diag}(\cdot)$ denote the statistical expectation and diagonalization operator, respectively.
$\lVert \mathbf{x} \rVert_{2}$ indicates the $2$-norm of a vector $\mathbf{x}$, and $\lVert \mathbf{A} \rVert_{F}$ stands for the Frobenius norm of a  matrix $\mathbf{A}$ with
arbitrary size and  $\odot$ stands for the Hadamard product (element-wise multiplication).

\begin{figure} 	
	\centering
	\includegraphics[scale=0.5]{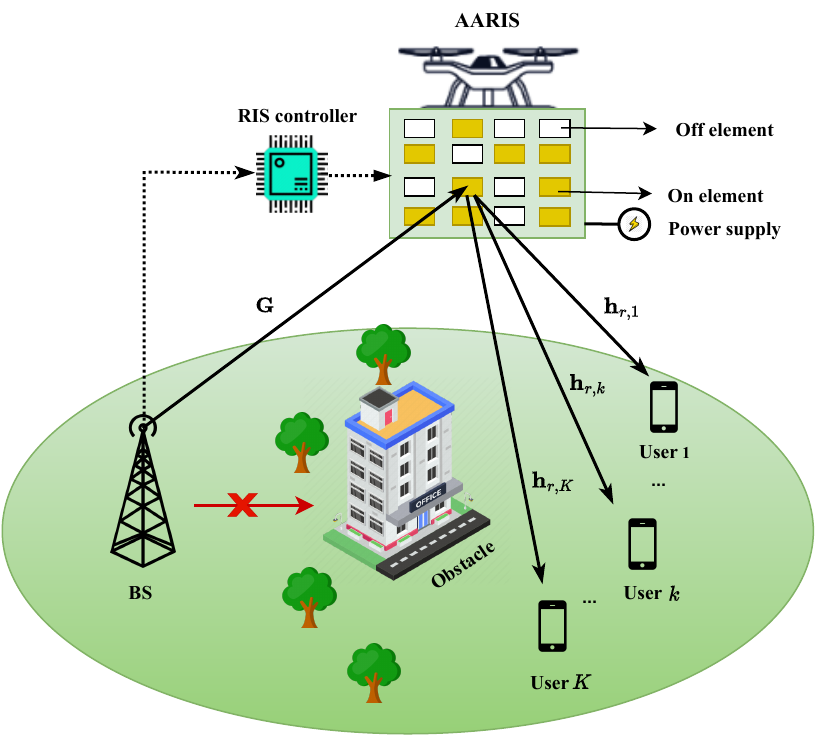}
	\caption{An AARIS system.}
	\label{system_model_1} 
\end{figure}

\section{System Model}\label{system model}

As depicted in Fig. \ref{system_model_1}, we consider a wireless cellular system, in which a BS serves multiple users in a donwlink transmission. Due to the existence of environmental obstructions, it is assumed that establishing LoS links for all users is not possible and the channel quality between the transceivers thus is not favourable enough to serve users effectively. With the assistance of an aerial communication, 
an active RIS is mounted over a rotary wing UAV, realizing an AARIS system, to reflect the transmitted signals from the BS to ground users. In this system, we assume that the BS is equipped with $N_{\text{BS}}$ antennas to serve a set $\mathcal{K} = \left\{ {1,2,...,K} \right\}$ of $K$ single-antenna users. A uniform planar array (UPA) with ${\mathcal{M}} = \left\{{{1},{2},...,{{{m_{{x}}}{m_{{y}}}}},...,{{M_{{x}}}{M_{{y}}}}} \right\}$ reflective units is deployed on the RIS, in which the number of reflecting
elements along the $x$- and $y$-axis are denoted by ${M_{{x}}}$ and ${M_{{y}}}$, respectively. Without loss of generality, a three-dimensional (3D) Cartesian coordinates models the locations of the BS, the AARIS, and the users. We consider a time interval $T$ to evaluate the performance of the network, consisting a set  $\mathcal{L}=\left\{ {1,2,...,L} \right\}$ of $L$ time slots, each with a duration of ${\tau}$, such that $T=L\tau$, and the network configuration is supposed to remain static within each slot. The BS has a fixed location, denoted by ${\bold{q}_\text{BS}}= {\left[ {x_{\text{BS}},y_{\text{BS}},z_{\text{BS}}} \right]^T}\in \mathbb{R}^{3 \times 1} $, whereas the location of users and the AARIS vary across consecutive time slots. The location of BS along the $x$-axis, $y$-axis, and $z$-axis are denoted by $x_{\text{BS}}$, $y_{\text{BS}}$, and $z_{\text{BS}}$, respectively. Moreover, the locations of user $k$ and the AARIS in time slot $l \in L$, $0 < l < L$, are defined as ${\bold{q}_k}\left(l \right) = {\left[ {x_{k}\left( l \right),y_{k}\left( l \right),0} \right]^T}\in \mathbb{R}^{3 \times 1} $, $\forall k \in \mathcal{K}$, and ${\bold{q}_{u}}\left( l \right) = {\left[ {x_{u}\left( l \right),y_{u}\left( l \right), H} \right]^T}\in \mathbb{R}^{3 \times 1} $, 
respectively, where $x_{k}\left( l \right)$ and $y_{k}\left( l \right)$ represent the location of $k$-th user in time slot $l$ along the $x$-axis and $y$-axis, respectively. Furthermore, the AARIS flies at the fixed altitude $H$, and the location of AARIS in time slot $l$ along the $x$-axis and $y$-axis are denoted by $x_{u}\left( l \right)$ and $y_{u}\left( l \right)$, respectively. In time slot $l$, the AARIS is flying with the velocity of ${\bold{v}_{u}}\left( l \right) = {\left[ {v_{u}^x\left( l \right),v_{u}^y\left( l \right), 0} \right]^T}\in \mathbb{R}^{3 \times 1}$, such that its initial position is $\mathbf{q}_{u}^{I}$. The velocity of AARIS in time slot $l$ along the $x$-axis and $y$-axis are defined as $v_{u}^x\left( l \right)$ and $v_{u}^y\left( l \right)$, respectively.
We model the flight of the AARIS by setting the following position and velocity constraints:
\begin{subequations}\label{eq5}
	\begin{align}
	& {\bold{q}_{u}}\left( {l + 1} \right) = {\bold{q}_{u}}\left( l \right) + {\bold{v}_{u}}\left( l \right){\tau}, 
	\quad \forall l \in \mathcal{L}, \label{c1}\\
	&{\bold{q}_{u}}\left[ 1 \right] = {\bold{q}_{u}^{\textrm{I}}}, \label{c2}\\
	&{\bold{q}_{\min }} \le {\bold{q}_{u}\left( l \right)} \le {\bold{q}_{\max }}, \quad \quad \forall l \in \mathcal{L}, \label{c3}\\
	&\left\| {{\bold{v}_{u}}\left( {l + 1} \right) - {\bold{v}_{u}}\left( l \right)} \right\| \le a _{\text{max}, u}{\tau}, \quad \forall l \in \mathcal{L}, \label{c4} \\
	&\left\| {{\bold{v}_{u}}\left( l \right)} \right\| \le V_{\text{max}, u}, \quad \quad \forall l \in \mathcal{L}, \label{c5}
    \end{align}
\end{subequations}
where \eqref{c1} determines its next location with respect to the movement velocity; \eqref{c2} shows the initial location ${\bold{q}_{u}^{\textrm{I}}}$ in the first time slot $l$; \eqref{c3} limits the UAV's movement to the minimum and maximum boundaries ${\bold{q}_{\min}}$ and ${\bold{q}_{\max }}$, respectively; finally, \eqref{c4} and \eqref{c5} control its movement acceleration and speed by introducing the maximum flight acceleration $a _{\text{max}, u}$ and the maximum flight speed $V _{\text{max}, u}$, respectively.
\vspace{-1.05em}
\subsection{Channel Model}
\vspace{0em}
\begin{figure}[ht] 	
	\centering
 \includegraphics[scale=0.5]{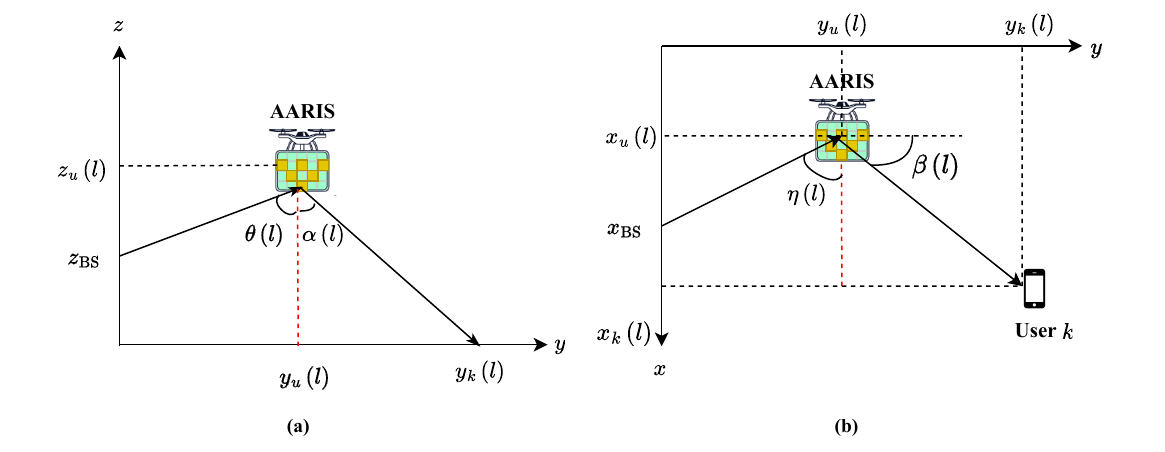}
	\caption{(a) and (b) show the vertical and horizontal angles between the BS and AARIS, respectively, as well as the AARIS and users in time slot $l$.}
	\label{angles_1_2} 
\end{figure}
\vspace{-0.2em}
We assume that the channel state information (CSI) is available at the BS via existing channel estimation schemes for each time slot $l$ \cite{CSI1}\footnote{ However, the existing channel estimation techniques can be well applied in the proposed resource allocation mechanism for more practical results, which resides beyond the scope of this paper.}. We adopt ${\bold{G}}(l) \in {\mathbb{C}^{M \times N_\text{BS}}}$ and  ${\bold{h}_{_{\textrm{r},k}}}(l) \in {\mathbb{C}^{{M} \times 1}}$, to denote the channel gain from the BS to the AARIS and from the AARIS to user $k$,  respectively. The direct LoS link between the BS and users blocked by environmental obstacles. 
The reflection coefficient of the active RIS is expressed as $\lambda_{m_x,m_y}\left( l \right)=a_{m_x,m_y}\left( l \right) e^{j\varphi_{m_x,m_y}}\left( l \right)$ in which $j=\sqrt{-1}$ represents the imaginary unit, $a_{m_x,m_y}\left( l \right) \in \mathbb{R}_{+}$ and $\varphi_{m_x,m_y}\left( l \right)$ $\in$ $(0,2\pi]$, $\forall m_x,m_y \in \mathcal M$ stand for the amplification factor and the phase shift of the (${{{m_{{x}}},{m_{{y}}}}}$)-th reflecting element of the active RIS in time slot $l$, respectively. Then, the diagonal amplification factor and the phase shift matrix for the active RIS in this time slot is given by
\begin{align}
\mathbf{\Lambda}(l)=\mathrm{diag}\left( \lambda_{1,1}\left( l \right), \dots, \lambda_{m_x,m_y}\left( l \right), \dots, \lambda_{M_x,M_y}\left( l \right) \right).
\end{align}
Accordingly, $\mathbf{\Lambda}(l) = \mathbf{A}(l) \mathbf{\Theta}(l)$, with the amplification factor matrix $\mathbf{A}(l) = \mathrm{diag}\left(  \mathbf{a}(l)\right) \in \mathbb{R}_{+}^{M \times M} $, and the phase shift matrix $\mathbf{\Theta}(l)=\mathrm{diag}\left(  \mathbf{\thetab}(l)\right) \in \mathbb{C}^{M \times M}$, such that 
\begin{align}
\mathbf{a}(l) = [a_{1,1}\left( l \right),\dots,a_{m_x,m_y}\left( l \right),\dots a_{M_x,M_y}\left( l \right)]^T,
\end{align}
and $\mathbf{\thetab}(l) = [e^{j\varphi_{1,1}}\left( l \right),\dots e^{j\varphi_{m_x,m_y}}\left( l \right), \dots, e^{j\varphi_{M_x,M_y}}\left( l \right)]^T$. Without loss of generality, we indicate ${m_x,m_y}$ and ${M_x,M_y}$, by $m$ and $M$, respectively. In the following, we will model the channel gain between the BS and AARIS, as well as between the AARIS and users, respectively.
\subsubsection{BS-AARIS channel}
Both LoS and non-line-of-sight (NLoS) links are taken into account for modelling the channel gain from the BS to the AARIS. Hence, on the basis of Rician fading, the channel gain between the BS and AARIS in time slot $l$ can be modelled as:

\vspace{-0.8em}
\small

\begin{align}
\begin{split}
{\bold{G}}(l) =& \sqrt {{C_0}{{\left( {\frac{{{d_{\text{BS},u}}(l)}}{{{D_0}}}} \right)}^{ - \alpha_{\text{BS},u}}}} \\
&\times\left\{ {\sqrt {\frac{K_{\text{BS},u}}{{{K_{\text{BS},u}} + 1}}} \bold{G}^{\text{LoS}}(l) + \sqrt {\frac{1}{{{K_{\text{BS},u}} + 1}}} \bold{G}^{\text{NLoS}}(l)} \right\},
\end{split}
\end{align}
\normalsize
where ${K_{\text{BS},u}}\ge 0$ and $\alpha_{\text{BS},u}\ge 0$ denote the Rician factor and the path loss exponent, respectively; $C_0$ and ${\bold{G}^{\text{NLoS}}(l)} \in \mathbb{C}^{M \times N_{\text{BS}}}$, respectively express the path loss at the reference distance $D_{0} = 1~m$ and the NLoS component of $\mathbf{G}(l)$; ${{{d_{\text{BS},u}}(l)}}$ represents the distance between the BS and AARIS, denoted by ${d_{\text{BS},u}} = {\left\| {{\bold{q}_{\text{BS}}}\left( l \right) - {\bold{q}_{u}}\left( l \right)} \right\|^2}$; $\bold{G}^{\text{LoS}}(l) \in \mathbb{C}^{M \times N_{\text{BS}}}$ and $\bold{G}^{\text{LoS}}(l)={\boldsymbol{\vartheta} _{\text{AoA},u}}(l)\times \boldsymbol{\vartheta} _{\text{AoD},\text{BS}}^T(l)$ with ${\boldsymbol{\vartheta}_{\text{AoA}, u}}(l) \in \mathbb{C}^{M \times 1}$ indicating, the angle of arrivals (AoAs), corresponding to the signal from the BS to the AARIS, as well as ${\boldsymbol{\vartheta} _{\text{AoD},\text{BS}}}(l) \in \mathbb{C}^{N_{\text{BS}} \times 1}$, denoting, the angle of departures (AoDs), corresponding to the signal from the BS to the AARIS, which can be modelled as
\begin{equation}
\begin{split}
{\boldsymbol{\vartheta}_{\text{AoA}, u}}(l) =& {\left[ {1,\dots,{e^{ - j\zeta ({M_{{x}}} - 1)\sin {\theta}(l)\cos {\eta}(l)}}} \right]^T}\\
&\otimes{\left[ {1,\dots,{e^{ - j\zeta ({M_{{y}}} - 1)\sin {\theta}(l)\cos {\eta}(l)}}} \right]^T},
 \end{split}
\end{equation}
\vspace{-0.5em}
and
\begin{align}
&{\boldsymbol{\vartheta} _{\text{AoD},\text{BS}}}(l)={\left[ {1,\dots,{e^{ - j\zeta ({N_{\text{BS}}} - 1)\sin {\theta}(l)\cos {\eta}(l)}}} \right]^T},
\end{align}
respectively. In the above equations, $\zeta  = \frac{{2\pi {f_c}d_{\textrm{RIS}}}}{c}$, with the speed of light $c$, the carrier frequency $f_c$, and the distance between two adjacent reflecting elements at the RIS, denoted by $d_{\textrm{RIS}}$.
Besides, for both AoD and AoA, the vertical and horizontal angles between the BS and AARIS in time slot $l$ are respectively denoted by ${\theta}(l)$ and ${\eta}(l)$, and defined as $\sin {\theta}(l) = \left( {\frac{{y_{u}(l)}}{{\sqrt {{{\left( {x_{u}(l)} \right)}^2} + {{\left( {y_{u}(l)} \right)}^2}} }}} \right)$ and $\cos {\eta}(l) = \left( {\frac{{x_{u}(l)}}{{\sqrt {{{\left( {x_{u}(l)} \right)}^2} + {{\left( {y_{u}(l)} \right)}^2}} }}} \right)$, as represented in Fig.2.
\vspace{0.8em}
\subsubsection{AARIS-user channel}
Similarly, the channel gain between the AARIS and user $k$ in time slot $l$ follows Rician fading and is denoted by ${\bold{h}_{{r},k}(l)}\in \mathbb{C}^{M \times 1}$. We model ${\bold{h}_{{r},k}(l)}$ as follows:
\vspace{-0.8em}

\small
\begin{equation}
\begin{split}
{\bold{h}_{{r},k}}(l) =& \sqrt {{C_0}{{\left( {\frac{{{d_{u,k}}(l)}}{{{D_0}}}} \right)}^{ - \alpha_{u,k}}}}\\
&\times\left\{ {\sqrt {\frac{K_{u,k}}{{{K_{u,k}} + 1}}} \bold{h}_{\textrm{r},k}^{\text{LoS}}(l) + \sqrt {\frac{1}{{{K_{u,k}} + 1}}} \bold{h}_{\textrm{r},k}^{\text{NLoS}}(l)} \right\},
\end{split}
\end{equation}
\normalsize
where ${\bold{h}_{{r},k}^{\text{NLoS}}(l)} \sim \mathcal{C}\mathcal{N}\left( {\boldsymbol{0},\bold{I}_{M}} \right) $, ${K_{u,k}}$, $\alpha_{u,k}$, and ${{{d_{u,k}}(l)}}$ specify the Rician factor, the path loss exponent, and the distance between the AARIS and user $k$ in time slot $l$, respectively, such that ${d_{u,k}}\left( l \right) = {\left\| {{\bold{q}_{{u}}}\left( l \right) - {\bold{q}_{k}}\left( l \right)} \right\|^2}$. The LoS channel gain ${\bold{h}_{{r},k}^{\text{LoS}}(l)} = {\boldsymbol{\vartheta}_{\text{AoD},u}}(l)$, where ${\boldsymbol{\vartheta}_{\text{AoD},u}}(l) \in \mathbb{C}^{M \times 1}$ denotes the AoDs from the AARIS to user $k$ and can be obtained as follows
\begin{align}
\begin{split}
{\boldsymbol{\vartheta}_{\text{AoD},u}}(l)=&{\left[ {1,\dots,{e^{ j\zeta ({M_{{x}}} - 1)\sin {\alpha}(l)\cos {\beta}(l)}}} \right]^T} \\ & \otimes {\left[ {1,\dots,{e^{ j\zeta ({M_{{y}}} - 1)\sin {\alpha}(l)\cos {\beta}(l)}}} \right]^T},
\end{split}
\end{align}
where $\sin {\alpha}(l) = \left( {\frac{{y_{k}(l) - y_{u}(l)}}{{\sqrt {{{\left( {z_{k}(l) - z_{u}(l)} \right)}^2} + {{\left( {y_{k}(l) - y_{u}(l)} \right)}^2}} }}} \right)$ and $\cos {\beta}(l) = \left( {\frac{{y_{k}(l) - y_{u}(l)}}{{\sqrt {{{\left( {x_{k}(l) - x_{u}(l)} \right)}^2} + {{\left( {y_{k}(l) - y_{u}(l)} \right)}^2}} }}} \right)$.

\normalsize

\subsection{Transmission Protocol}
The BS leverages RSMA as an efficient multiple access scheme, through which each user message is splitted into common and private parts before the transmission. Whereas the common message contains an identical content for all users, each user has its specific private message with a unique content. By integrating the
messages for $K$ users into one superimposed signal, there are totally $K+1$ messages, i.e., one common message for all users, in addition to $K$ private messages, each corresponds to one user.
In light of the above description, we define a common stream $s_{\textrm{c}} \in \mathbb{C}$ for all users, as well as $K$ private streams ${s_{k}}\in \mathbb{C}, ~\forall k \in \mathcal{K}$. Accordingly, in time slot $l$, we define a transmit beamforming vector associated with the common message, denoted by ${{\bold{w}_{\textrm{c}}}} (l) \in \mathbb{C}^{N_{\text{BS}} \times 1}$, as well as $K$ transmit beamforming vectors for $K$ private messages, denoted by ${{\bold{w}_k}} (l) \in \mathbb{C}^{N_{\text{BS}} \times 1}$. Without loss of generality, it is assumed that $\mathbb{E}\lbrace \lvert s_{\textrm{c}} \rvert^{2} \rbrace = 1$, $\mathbb{E}\lbrace \lvert s_{k} \rvert^{2} \rbrace = 1,~ \mathbb{E}\lbrace {s}_{c}^* s_{k} \rbrace = 0, ~\forall k \in \mathcal{K}$. Then the superimposed transmit signal can be expressed as  
\vspace{-0.1em}
\begin{align}
	\bold{x}(l) = \sum\limits_{k = 1}^K {{\bold{w}_k}} (l){s_k}+{{\bold{w}_{\textrm{c}}}} (l){s_{\textrm{c}}}.
\end{align}
As such, the received signal at the $k$-th user in time slot $l$ can be expressed as
\begin{equation} \label{eq10}
\begin{split}
{y_k}\left( l \right) &= 
 \bold{h}_{\textrm{r},k}^H(l){\bold{\Lambda}}(l) \Big(\mathbf{G}(l) \mathbf{x}(l) + \mathbf{z}\Big)  + n_{k},\\
\end{split}
\end{equation}
where $\mathbf{z} \sim \mathcal{C}\mathcal{N} (\mathbf{0}_{M},\sigma _{z}^2 \mathbf{I}_{M})$ and ${n_k} \sim \mathcal{C}\mathcal{N}\left( {0,\sigma _k^2} \right)$ denote the dynamic noise generated by the AARIS \cite{Ref1} and the additive white Gaussian noise (AWGN) of the $k$-th user with zero means and variances $\sigma _z^2$ and $\sigma _k^2$, respectively.\\
Depending on the channel condition, each element of the active RIS can be adaptively switched off to enhance the overall system EE. To this end, the binary element selection matrix $\mathbf{F}(l) = \text{diag}(\mathbf{f(l)}) \in \lbrace 0, 1 \rbrace^{M \times M}$ is introduced, such that $\mathbf{f}(l) = [f_{1}(l), \dots, f_{M}(l)]^T \in \lbrace 0, 1 \rbrace^{M \times 1}$. More specifically, $f_{i}(l)=1$ denotes that the $i$-th element of the active RIS is on in time slot $l$ and $f_{i}(l)=0$, otherwise \cite{Ref3}. 
By taking the element selection matrix into consideration, \eqref{eq10} could be rewritten as:
\begin{equation} \label{eqhada}
\begin{split}
{y_k}\left( l \right) &= 
\bold{h}_{\textrm{r},k}^H(l)\Big(\mathbf{\Lambda}(l) \odot \mathbf{F}(l) \Big) \Big(\mathbf{G}(l) \mathbf{x}(l) + \mathbf{z}\Big)  + n_{k}.
\end{split}
\end{equation}
Let $\mathbf{F^\prime}(l)=\Big(\mathbf{\Lambda}(l) \odot \mathbf{F}(l) \Big)$, $\mathbf{h}_{{r},k}^H(l) \in \mathbb{C}^{1 \times N_{\text{BS}}}$ and $\mathbf{h}_{k}^H(l) = \bold{h}_{{r},k}^H(l) \mathbf{F^\prime}(l) \mathbf{G}(l)$. Then, \eqref{eqhada} can be further rewritten as follows:
\begin{equation} \label{eq11}
\begin{split}
{y_k}\left( l \right) &=  \bold{h}_{{r},k}^H(l) \mathbf{F^\prime}(l) \mathbf{G}(l) \mathbf{x}(l) + \bold{h}_{{r},k}^H(l) \mathbf{F^\prime}(l) \mathbf{z} + n_{k},\\
&= \mathbf{h}_{k}^H(l) \mathbf{x}(l) + \bold{h}_{{r},k}^H(l) \mathbf{F^\prime}(l) \mathbf{z} + n_{k}.
\end{split}
\end{equation}
\vspace{0.1em}
\!\!\!On the other hand, to exploit RSMA, the common stream is decoded by each user in time slot $l$, while assuming the private streams as interference in the proposed system model \cite{Ref4, Ref5}. Hence, the received signal-to-interference-plus-noise ratio (SINR) of the common
stream at the $k$-th user in time slot $l$ can be expressed as
\vspace*{-1em}

\begin{small}
\begin{equation}
\begin{split}
& {\gamma _{\textrm{c},k}}(l)  =\frac{ \lvert \bold{h}_k^H(l){\bold{w}_c}(l)  \rvert^{2} }{ \sum\limits_{i = 1}^K \lvert \bold{h}_k^H(l){\bold{w}_i}(l) \rvert^{2} + \sigma_{z}^{2} \lVert \mathbf{h}_{\textrm{r},k}^{H}(l) \mathbf{F^\prime}(l) \rVert^{2} + \sigma_{k}^{2}}, \ \forall k \in \mathcal{K}.
\end{split}
\end{equation}
\end{small}
\normalsize
\vspace{0.01em}
\!\!More specifically, after decoding the common stream, each user applies successive interference cancellation (SIC) to decode its own private stream by eliminating the influence of the common stream. Then, the private SINR of the $k$-th user is given by
\vspace*{-1em}

\small
\begin{equation}
\begin{split}
& {\gamma _{\textrm{p},k}}(l) = \frac{ \lvert \bold{h}_k^H(l){\bold{w}_k}(l)  \rvert^{2} }{ \sum\limits_{i = 1,i \ne k}^K \lvert \bold{h}_k^H(l){\bold{w}_i}(l) \rvert^{2} + \sigma_{z}^{2} \lVert \mathbf{h}_{\textrm{r},k}^{H}(l) \mathbf{F^\prime}(l) \rVert^{2} + \sigma_{k}^{2}}, \ \forall k \in \mathcal{K}. 
\end{split}
\end{equation}
\normalsize
\vspace{-0.1em}	
\!\!On the basis of Shannon-Hartley theorem \cite{Ref7}, the achievable rate information in bps/Hz corresponded to the common and private stream at the $k$-th user in time slot $l$ can be respectively expressed as
\begin{equation} \label{rate}
\begin{split}
& R_{\textrm{c},k}(l) = \log_{_2}(1 + \gamma_{\textrm{c},k}(l)), \ \ \forall k \in \mathcal{K}, \\ 
& R_{\textrm{p},k}(l) = \log_{_2}(1 + \gamma_{\textrm{p},k}(l)), \ \ \forall k \in \mathcal{K}.
\end{split}
\end{equation}
To unlock the full potential of RSMA, all users should be able to decode the common stream. In other words, the achievable data rate of the common stream could not exceed the minimum achievable data rate among all users in time slot $l$, i.e., $R_{\textrm{c}}(l) \leq \min \lbrace R_{\textrm{c},1}(l), R_{\textrm{c},2}(l), \dots, R_{\textrm{c},k}(l) \rbrace$. Assume that the common data rate $R_{\textrm{c}}(l)$ is shared by all $K$ users and let $C_{k}(l)$ denote
the portion of $R_{\textrm{c}}(l)$ transmitting the common stream. Then, $R_{\textrm{c}}(l) = \sum_{k=1}^{K} C_{k}(l)$ and
\begin{equation}
\sum_{k=1}^{K} C_{k}(l) \leq \min_k \lbrace R_{\textrm{c},k}(l)\rbrace, \forall k \in \mathcal{K}.
\end{equation}
As a result, the total achievable data rate of the system in time slot $l$ is given by
\begin{equation}
R_{\text{total}}(l) = \sum_{k = 1}^{K} \bigg(C_{k}(l) + R_{\textrm{p},k}(l) \bigg).
\end{equation}
\vspace{-3em}	
\subsection{Power Consumption}
The overall system power consumption consisting of the power consumption of the UAV and RIS is defined as follows.
\subsubsection{Power Consumption of UAV}
{The propulsion power consumption of rotary-wing UAV which is necessary to support its movement and keep it aloft for aerial operation can be mathematically modeled as \cite{Rotary_wing_UAV}
\vspace{-0.3em}	
	
\small	
\begin{equation} \label{P1}
\begin{split}
{{\mathop{\rm P}\nolimits} _{\text{UAV}}}\left( l \right) =&  \underbrace {{P_b}\left( {1 + \frac{{3 \lVert \vb_{u}(l) \rVert^{2}    }}{  \Omega_{u}^{2}  R_{u}^{2}    }} \right)}_{\text{blade profile}} + \underbrace {\frac{1}{2}{d_u}{\zeta \delta }{A_u}\lVert \vb_{u}(l) \rVert^{3} }_{\text{parasite}} \\
& + \underbrace {{P_i}\left( {\sqrt {\sqrt {1 + \frac{{\lVert \vb_{u}(l) \rVert^{4} }}{{4v_{u,i}^4}}}  - \frac{{\lVert \vb_{u}(l) \rVert^{2} }}{{2v_{u,i}^2}}} } \right)}_{\text{induced}}
,
\end{split}
\end{equation}
\normalsize
where the blade angular velocity in Radian per second (rad/s) and the UAV rotor radius in meter (m) are denoted by $\Omega_{u}$ and $R_u$, respectively, while the average rotor induced velocity is $v_{u,i}$. Moreover, ${d_u}$, $\zeta$, $\delta$, and ${A_u}$ denote the fuselage drag ratio, air density, rotor solidity, and rotor disk area of the UAV, respectively.  The blade profile power and induced power in hovering mode are represented by two constants, namely, ${P_b}$ and ${P_i}$, respectively. Then, the power consumption for hovering is derived by substituting $v_{u}\left( l \right)=0$ into $\eqref{P1}$ that yields \cite{Rotary_wing_UAV}
\begin{align} \label{P2} 
{P_{u,h}} = \underbrace {\frac{\rho }{8}\zeta \delta {A_u} \Omega_{u}^{3} R_u^3}_{\buildrel \Delta \over = {P_b}} + \underbrace {\left( {1 + \iota } \right)\frac{{W_u^{3/2}}}{{\sqrt {2\zeta {A_u}} }}}_{ \buildrel \Delta \over = {P_i}},
\end{align} 
where $\rho\ge0$ is a profile drag coefficient, while  $\iota\ge0$ is an incremental correction factor in the induced power and $W_u$ represents the total UAV weight in Newton.
\subsubsection{Power Consumption of RIS}{
The required instantaneous power for the RIS in time slot $l$ is given by
\begin{align} \label{p3}
P_{\text{RIS}}(l) = N(P_{\text{c}} + P_{\text{DC}}) + \nu p_{\text{out}}(l),
\end{align}
where $\nu \triangleq \eta^{-1}$ with $0 \le \eta \le 1$ being the amplifier efficiency, while $P_{\text{c}}$ and $P_{\text{DC}}$ denote the constant power consumption of the switching and control 
circuits and biasing power consumption at each reflecting element, respectively. Moreover, $p_{\text{out}}(l)$ is the output power of the active RIS that can be expressed as  
\begin{align} \label{p4}
\nonumber{p_{\text{out}}}(l)= &\sum\limits_{k = 1}^K {{{\left\|{\mathbf{F^\prime}(l)  {\bold{G}(l)}{\bold{w}_{\textrm{p},k}}\left( l \right)} \right\|}^2}}+{{{\left\|{\mathbf{F^\prime}(l)  {\bold{G}(l)}{\bold{w}_c}\left( l \right)} \right\|}^2}}\\ &+\sigma _z^2{\left\| {\mathbf{F^\prime}(l)} \right\|^2_F}.
\end{align}}
\vspace*{-0.5em}
\subsubsection{Total Power Consumption}{Based on \eqref{p3} and \eqref{p4}, the total power consumption of the system is given by
\begin{equation}\label{t_power}
 P_{\text{total}}(l) = 1/\epsilon \sum_{k=1}^{K} \lVert \mathbf{w}_{k}(l) \rVert^{2} +1/\epsilon \lVert\mathbf{w}_{\textrm{c}}(l) \rVert^{2}+ P_{\text{Cir}} + P_{\text{AARIS}}(l),
\end{equation}
where $0<\epsilon \le 1$ is the power amplifier efficiency at the BS, and $P_{\text{Cir}} = P_{\text{BS}}^{\text{Cir}} + \sum_{k=1}^{K} p_{k}^{\text{Cir}}$ denotes a constant circuit power dissipation including the static power consumption of the BS ($P_{\text{BS}}^{\text{Cir}}$) and users ($p_{k}^{\text{Cir}}$). Furthermore, the power consumption of AARIS which is the aggregation of the UAV propulsion power consumption and RIS power consumption in time slot $l$ that can be expressed as 
\begin{align} \label{p-tot}
P_{\text{AARIS}}(l) = P_{\text{UAV}}(l) + P_{\text{RIS}}(l).
\end{align}
}
}
\section{Problem Formulation}
In this section, we formulate an optimization problem to maximize the overall system EE in the discussed RSMA-enabled AARIS network, by jointly designing the active beamforming $\mathbf{W} = \lbrace \mathbf{w}_{k}(t), \forall  k, t\rbrace$ matrix, UAV joint trajectory and velocity $\mathbf{q} = \left\{ {{\bold{q}_{{u}}}(t), {{\bold{v}_{{u}}}(t)}, \forall t} \right\}$, the active RIS amplification factor, and phase shift matrix $\mathbf{\Lambda} = \left\{ {{\bold{\Lambda}}(t), \forall t} \right\}$, data rate for receiving the common message $\mathbf{c} = \lbrace C_{k}(t), \forall k, t   \rbrace$, and $\mathbf{F} = \lbrace \mathbf{F}(t), \forall t \rbrace $ for element selection at the RIS. Furthermore, the EE of the entire system is given by 
\begin{equation} \label{EE}
\text{EE}(\mathbf{W}, \bold{q}, \bold{\Lambda}, \bold{F}, {\bold{c}}) = \frac{1}{L} \sum_{t=0}^{L} \frac{R_{\text{total}}(l)}{P_{\text{total}}(l)}.
\end{equation} 
Consequently, the design can be formulated as the following optimization problem.
\begin{equation} 
\begin{split}
{\mathcal{P}_0}: &\mathop {\text{max}}\limits_{\mathbf{W}, \bold{Q}, \bold{\Lambda},\bold{F}, {\bold{C}}} \text{EE}(\mathbf{W}, \bold{Q}, \bold{\Lambda}, \bold{F}, {\bold{C}}),\nonumber \\
\textrm{s.t.} \quad & \text{C}_1:\sum_{k=1}^{K} C_{k}(l) \leq \min_k \lbrace R_{\textrm{c},k}(l)\rbrace, \quad \forall k \in \mathcal{K}, \forall l \in \mathcal{L}, \nonumber \\
\quad& \text{C}_{2}: C_{k}(l) + R_{\textrm{p},k}(l) \geq \Pi_{k}\textcolor{black}{(l)}, \  \quad \forall k \in \mathcal{K}, \forall l \in \mathcal{L}, \nonumber \\
\quad& \text{C}_3:1/\epsilon\sum\limits_{k = 1}^K {{{\left\| {{\bold{w}_k}(l)} \right\|}^2}}+1/\epsilon{{{\left\| {{\bold{w}_c}(l)} \right\|}^2}}  \le P_{\text{max},\text{BS}}, \forall l  \in \mathcal{L}, \nonumber\\
\quad& \text{C}_4: {p_{\text{out}}}(l) \le \eta P_{\text{I}}, \quad \quad \forall l  \in \mathcal{L}, \nonumber\\ 
\quad& \text{C}_5:0 \le a_{m}(l) \le a_{\max}, \quad \quad \forall l  \in \mathcal{L}, \forall m \in \mathcal{M},  \nonumber\\
\quad& \text{C}_6: 0 \le {\varphi}_{m}(l) \le 2\pi, \quad \quad \forall l \in \mathcal{L}, \forall m \in \mathcal{M}, \nonumber\\
\quad & \text{C}_7: {\bold{q}_{\min }} \le {\bold{q}_{u}\left( l \right)} \le {\bold{q}_{\max }}, \quad \quad \forall l \in \mathcal{L}, \nonumber\\
\quad& \text{C}_8: {\bold{q}_{u}}\left( {t + 1} \right) = {\bold{q}_{u}}\left( l \right) + {\bold{v}_{u}}\left( l \right){\tau}, \quad \forall l \in \mathcal{L},  \nonumber\\
\end{split}
\end{equation}
\begin{align}
\quad& \text{C}_{9}: {\bold{q}_{u}}\left[ 1 \right] = {\bold{q}_{u}^{I}}, \nonumber\\
\quad& \text{C}_{10}: \left\| {{\bold{v}_{u}}\left( l \right)} \right\| \le V_{\text{max}, u}, \quad \quad \forall l \in \mathcal{L},  \nonumber\\
\quad& \text{C}_{11}: \left\| {{\bold{v}_{u}}\left( {t + 1} \right) - {\bold{v}_{u}}\left( l \right)} \right\| \le a _{\text{max}, u}{\tau}, \quad \forall l \in \mathcal{L},  \nonumber\\
\quad& \text{C}_{12}: \sum_{m=1}^{M} f_{m}(l) \leq M, \quad \forall l \in \mathcal{L}, \nonumber\\
\quad& \text{C}_{13}: f_{m}(l) \in \lbrace 0,1 \rbrace, \quad \forall l \in \mathcal{L}, m \in \mathcal{M},
\end{align}
\noindent where constraint $\text{C}_1$ ensures that the RSMA common message is able to be successfully decoded at all users in time slot $l$; constraint $\text{C}_2$ guarantees the QoS threshold $\Pi_{k}(l)$ for the $k$-th user in time slot $l$; In $\text{C}_3$ and $\text{C}_4$, the transmit power of the BS and amplification power budget for RIS in time slot $l$ are limited to $P_{\text{max},\text{BS}}$ and $P_{\text{I}}$, respectively. The reflection coefficient of the active RIS is bounded to $a_{\textrm{max}}$ in $\text{C}_5$; constraint $\text{C}_6$ limits the reflection of each RIS element $m$ in time slot $l$ within $[0,2\pi]$. The flight of the AARIS is controlled via $\text{C}_7$-$\text{C}_{11}$, elaborated under (\ref{eq5}) in previous section. We ensure that mention element selection constraint, in which the maximum $M$ on elements is considered in $\text{C}_{12}$ and finally, $\text{C}_{13}$ indicates the domain of the RIS element binary selection variables.
\par Regarding the fractional form of the objective function and non-convex constraints, this problem is non-convex. Furthermore, the coexistence of discrete and continuous optimization variables, makes this problem being in a mixed integer and non-linear programming (MINLP) form and thus is non-deterministic polynomial-time hardness (NP-hard) \cite{Hosein2,Hosein-G}. Therefore, obtaining a globally optimal solution to this problem via the brute-force method, e.g. an exhaustive search, is generally intractable even for a moderate size of system. Indeed, prior studies based on convex optimization techniques propose complex mathematical transformations so as to obtain a locally optimal solution. Moreover, due to the dynamic nature of wireless environment, a real-time adaptive is preferred. The recent DRL methods not only achieve this, but also propose a joint solution without decomposing the problem, as regarded in prior convex optimization methods. We also capture the dynamicity of the wireless environment so as to promote the efficiency of the DRL, by initializing its training parameters via meta-learning, which leads to better generalization and adaptability to system dynamics. Specifically, we devise a DRL method with modified-SAC and TD3  algorithms (termed MSAT algorithm in the sequel). In practice, the former optimizes the RIS element selection matrix $\mathbf{F}$ dealing with the discrete domain, whereas the latter jointly optimizes other optimization variables in the continuous domain. Finally, the proposed MSAT algorithm is fine-tuned by optimizing its training parameters via meta-learning.

\vspace*{-0.5em}
\section{Proposed DRL-Based Approach}
\vspace*{-0.2em}

\textcolor{black}{Model-free RL is a dynamic programming tool that solves optimization problems in a dynamic environment by acquiring an effective solution through learning. Specifically, in RL algorithms, the agent interacts with the environment, selects actions to maximize cumulative rewards including state, action, reward, and next state. Therefore, we formulate the system EE maximization problem for the AARIS system model as an RL problem and solve it employing MSAT algorithm. However, the AARIS system is highly dynamic due to the mobility of the UAV and users and the conventional RL-based solutions cannot cope with such a degree of mobility in an effective manner~\cite{Meta1, Meta_Faramarzi}.} 
\textcolor{black}{Recently, meta-learning has been introduced~\cite{MAML} for improving the limited generalization of conventional RL method, by training its learning agent in various scenarios and task. Inspired by this fact, we introduce a novel algorithm, namely MMSAT, which improves our first proposed MSAT algorithm by equipping with a model-agnostic meta-learning (MAML) approach~\cite{MAML}. The newly proposed MMSAT algorithm exhibits better adaptability and predictivity in upcoming unkown environments.}
RL methods introduce an agent that frequently interacts with the dynamic environment it operates in, and learns how to optimize it based on a predefined criterion. In this article, we consider a central controller at the BS as the RL agent, which interacts with the AARIS system (including the UAV-borne RIS, the BS and the users) as the RL environment. However, in order to exploit the real-time decision making capability of the RL method and propose a solution framework to optimization problem $\mathcal{P}_{0}$, it is required first to reformulate this problem in a Markov decision process (MDP) form.\\
\vspace*{-2em}
\subsection{Reinforcement Learning (RL)}

There are five key elements to define an RL problem:
\begin{itemize}
	\item \textbf{Action space:}\textcolor{black}{\ All the variables or choices available to an agent at any time $l$, which can be selected through interaction with the environment, are denoted as actions $\ab_{l}$. The set of all actions at all time slots is referred to as the action space $\mathcal{A}$, $ \mathbf{a}_{l} \in \mathcal{A}$. The action space can consist of discrete or continuous actions. In the context of an optimization problem formulated as an RL problem, the optimization variables correspond to the actions that the agent aims to select to maximize a specified reward function. In the optimization problem $\mathcal{P}_{0}$, the actions involve selecting the active beamforming vector $\lbrace \mathbf{w}_{k}(l), \forall l\rbrace_{k \in \mathcal{K}}$, the UAV joint trajectory and velocity $\left\{ {{\bold{q}_{{u}}}(l), {{\bold{v}_{{u}}}(l)}, \forall l} \right\}$, the diagonal amplification factor and phase shift matrix for AARIS $\left\{ {{\bold{\Lambda}}(l), \forall l} \right\}$, the common message data rate $\lbrace C_{k}(l), \forall l   \rbrace_{k \in \mathcal{K}}$, and the element selection matrix of AARIS $\lbrace \mathbf{F}(l), \forall l \rbrace $ to maximize the reward function. Notably, among these actions, $\lbrace \mathbf{F}(l), \forall l \rbrace $ is discrete, while the others are continuous. Therefore, the actions are categorized into two groups: discrete $\mathbf{a}_{l}^{(1)} = 	\mathbf{F}(l)$ and continuous $\mathbf{a}_{l}^{(2)} = \big \lbrace \lbrace \mathbf{w}_{k}(l) \rbrace_{k \in \mathcal{K}}, {\bold{q}_{{u}}(l), {{\bold{v}_{{u}}}(l)}} ,  {{\bold{\Lambda}}(l)}, \lbrace C_{k}(l)\rbrace_{k \in \mathcal{K}} \big \rbrace$. Accordingly, an action $\mathbf{a}_{l} \in \mathcal{A}$ is defined as $\mathbf{a}_{l} = \lbrace \mathbf{a}_{l}^{(1)}, \mathbf{a}_{l}^{(2)} \rbrace$ in time slot $l$, and the action space is $\mathcal{A} = \lbrace \mathbf{a}_{l}, \forall l \rbrace$. For handling continuous actions, the TD3 algorithm is employed, while the modified-SAC algorithm is utilized for discrete actions.}\\
	\item \textbf{State space:}\textcolor{black}{\ The information that the agent receives from the environment at any time slot $l$ related to problem $\mathcal{P}_{0}$ is called state $\mathbf{s}_{l}$. The set of all states at all time slots is called state space $\mathcal{S} = \lbrace \mathbf{s}_{l} \rbrace_{l \in \mathcal{L}}$, where  $\mathbf{s}_{l} \in \mathcal{S}$. 
		All the QoSs mentioned in the constraints are considered part of the state. The state is defined to include the achievable common and private rates $\lbrace R_{c,k}(l), R_{p,k}(l) \rbrace_{\forall k \in \mathcal{K}}$ for all users, the maximum power of BS $P_{\text{max},\text{BS}}$, maximum amplification factor of each RIS element $a_{\max}$, location and speed of UAV at any time slot $l$ ${\bold{q}}_{u}\left( l \right), {\bold{v}_{u}}\left( l \right)$, location of all users $\lbrace \mathbf{q}_{k}(l) \rbrace_{k \in \mathcal{K}}$, maximum attainable acceleration of the UAV $a _{\text{max}, u}$, channel coefficients between the BS and AARIS, as well as channel coefficients between the AARIS and all users. Since the channel coefficients are complex numbers, they cannot be given directly to the DNN; for this reason, each complex number is divided into real and imaginary parts. In addition to the information received from the environment, actions and rewards can help the agent to learn helpful information. Therefore, states can be divided into three parts: environmental information, action, and reward. The environmental information in time slot $l$ is denoted as $\mathbf{I}_{l}$ and defined as follows
		\begin{equation}
		\begin{split}
		\mathbf{I}_{l} & = \big \lbrace \lbrace R_{c,k}(l), R_{p,k}(l), \Pi_{k}(l) \rbrace_{\forall k \in \mathcal{K}}, P_{\text{max},\text{BS}}, \mathbf{q}_{u}(l), \mathbf{v}_{u}(l), \\
		& \lbrace \mathbf{q}_{k}(l) \rbrace_{k \in \mathcal{K}}, a_{\max, u}, \mathbf{G}_{\text{real}}(l), \mathbf{G}_{\text{img}}(l),
		\lbrace \mathbf{h}_{r, k, \text{real}}(l)\rbrace_{k \in \mathcal{K}}, \\ & \lbrace \mathbf{h}_{r, k, \text{img}}(l)\rbrace_{k \in \mathcal{K}} \big \rbrace,
		\end{split}
		\end{equation}
		where $\mathbf{G}_{\text{real}}(l)$ is the real part of $\mathbf{G}(l)$ and $\mathbf{G}_{\text{img}}(l)$ is the imaginary part of $\mathbf{G}(l)$. Moreover, $	\lbrace \mathbf{h}_{r, k, \text{real}}(l)\rbrace_{k \in \mathcal{K}}$ denotes the real part of $	\lbrace \mathbf{h}_{r, k}(l)\rbrace_{k \in \mathcal{K}}$, while $\lbrace \mathbf{h}_{r, k, \text{img}}(l)\rbrace_{k \in \mathcal{K}} \big \rbrace$ represents the imaginary part of $\lbrace \mathbf{h}_{r, k}(l)\rbrace_{k \in \mathcal{K}} \big \rbrace$.
		Finally, the state $s_{l} \in \mathcal{S}$ is defined as follows
		\begin{equation}
		\mathbf{s}_{l} = \big \lbrace \mathbf{I}_{l}, \mathbf{a}_{l}, r(\mathbf{s}_{l}, \mathbf{a}_{l}) \big \rbrace,
		\end{equation}
		where $r(\mathbf{s}_{l}, \mathbf{a}_{l})$ represents the immediate reward in time slot $l$.} \\
	\item \textbf{Policy:} 
	\textcolor{black}{This paper utilizes two action selection policies, denoted as $\mu$ and $\pi$, respectively representing deterministic and stochastic policies, respectively. In a deterministic policy, states are mapped to actions, specifying the action selected in each state as $\mathbf{a}_{l} = \mu(\mathbf{s}_{l}) \in \mathcal{A}$. In stochastic policy, the probability of which action is chosen in each state is prescribed by determining the probability distribution function $\pi(\mathbf{a}_{l}|\mathbf{s}_{l})$, where $\pi(\mathbf{a}_{l}|\mathbf{s}_{l})$ is the probability of taking action $\mathbf{a}_{l}$ in state $\mathbf{s}_{l}$.}\\
	\item \textbf{Reward function:} 
	\textcolor{black}{Selecting an appropriate reward function is crucial as it serves as a signal indicating the suitability of the chosen action in the current state; this function is designed to maximize the EE objective function and satisfy the constraints in problem $\mathcal{P}_{0}$. Accordingly, the reward function is formulated as follows	
		\begin{equation}\label{reward_func}
		\begin{split}
		r_{l}(\mathbf{s}_{l}, \mathbf{a}_{l}) = \text{EE} + \sum_{i=1}^{13} \big(\beta_{i} \times \text{EE}\big),
		\end{split}
		\end{equation}
		where $\beta_{i}=0, \ i \in \lbrace 1, \dots, 13 \rbrace$ corresponds to $\text{C}_{1}$-$\text{C}_{13}$, such that if the constraint $\text{C}_{i}$
		is satisfied, and $-1$, otherwise. 
		As evident from \eqref{reward_func}, the reward function is designed such that no penalty is incurred if all constraints are satisfied, and vice versa. 
		\\Due to the continuous state space and the mix of continuous and discrete (binary) action spaces in problem $\mathcal{P}_{0}$, employing a $q$-table and the $q$-learning algorithm for its resolution is infeasible. On the other hand, DNNs, serving as parametric estimators, can be integrated with RL algorithms to handle high-dimensional continuous or discrete states and actions. In the realm of DRL, the objective is to design the optimal policy; this implies that the DNNs employed in the DRL algorithm are trained with the help of the training data in the replay buffer, consisting of $4$-tuples $(\mathbf{s}_{l}, \mathbf{a}_{l}, r(\mathbf{s}_{l}, \mathbf{a}_{l}), \mathbf{s}_{l+1})$. The parameters of the DNNs are set to map each received state $\mathbf{s}_{l}$ to an action $\mathbf{a}_{l}$ in the case of a deterministic policy or to determine the probability of selecting action $\mathbf{a}_{l}$ in the case of a stochastic policy.}\\ 
	\textcolor{black}{The majority of DRL algorithms are designed for continuous state and action spaces. However, the DQN algorithm receives the continuous state and selects a discrete action from the action space which has the highest $q$-value as an action. The action $\mathbf{a}_{l}^{(1)} \in \mathcal{A}$ consists of a binary vector of length $M$ (representing the number of AARIS elements); some of its arrays are one (on), and the rest are zero (off). Therefore, the DQN algorithm treats each element of this vector as two separate actions; one of the actions for showing the off RIS element and another for on RIS element. This necessitates the exploitation of $M$ distinct DQN networks.
	\\Converging and training $M$ DQN networks, however, yields sub-optimal results due to the high computational calculations.  Therefore, we propose the utilization of the modified-SAC algorithm. Indeed, SAC is an established DRL algorithm designed for continuous state and action spaces, and through specific modifications, we adapt it to handle discrete action spaces, with further details provided later. Two distinct algorithms are required to handle $\mathbf{a}_{l}^{(1)}$ and the other $\mathbf{a}_{l}^{(2)}$ action space. The MSAT algorithm operates in such a way that receives the state $\mathbf{s}_{l}$ at the input of both algorithms and gives action $\mathbf{a}_{l}^{(1)}$ at the output of the modified-SAC algorithm and action $\mathbf{a}_{l}^{(2)}$ at the output of the TD3 algorithm. The subsequent sections elaborate two modified-SAC and TD3 algorithms as follows}
\end{itemize}
\section{Modified SAC Algorithm}
\textcolor{black}{SAC is a model-free DRL algorithm that optimizes a stochastic policy using an off-policy approach \cite{SAC_algorithm}. The objective of the policy training is to maximize a trade-off between the expected return and entropy, representing the level of randomness in the policy. The state-action value function quantifies the performance of an agent in executing a particular action in a given state under a policy $\pi$, and it is defined as follows: 
\begin{equation}\label{state-action}
	\begin{split}
		q_{\pi}(\mathbf{s}_{l}, \mathbf{a}_{l}^{(1)}) =& \mathbb{E}_{(\mathbf{s}_{l}, \mathbf{a}_{l}^{(1)}) \sim P} \Big[\sum_{l=0}^{\infty} \gamma^{l} r(\mathbf{s}_{l}, \mathbf{a}_{l}^{(1)}) \\
		& - \lambda \sum_{l=1}^{\infty} \gamma^{l} \log_{10} \big( \pi(\cdot|\mathbf{s}_{l}) \big) \Big |
		\mathbf{s}_{l} = \mathbf{s}_{0}, \mathbf{a}_{l}^{(1)} = \mathbf{a}_{0}\Big],
	\end{split}
\end{equation}
where $P$ denotes the transition probability and $\gamma \in (0, 1]$ represents the discount factor. Moreover, the temperature parameter $\lambda$ determines the relative importance of the entropy term in comparison to the reward. Therefore, it determines the stochasticity of the optimal policy. The optimal policy is defined as follows:
\begin{equation}
	\pi^{\star} = \mathrm{arg} \ \underset{\pi}{\mathrm{max}} \quad q_{\pi}(\mathbf{s}_{l}, \mathbf{a}_{l}^{(1)}).
\end{equation}
The SAC algorithm is implemented adopting five DNNs: a critic network $1$ (CN1) with parameter $\thetab_{1}$, a critic network $2$ (CN2) with parameter $\thetab_{2}$, a target critic network $1$ (TCN1) with parameter $\bar{\thetab}_{1}$, a target critic network $2$ (TCN2) with parameter $\bar{\thetab}_{2}$, and the actor network (AN) with parameter $\phib$. Furthermore, CN$1$, CN$2$, TCN$1$, and TCN$2$ share the same structure. A new vector is obtained by concatenating the state vector and action vector together, with dimensions equal to the sum of the state and action vectors. 
\\The training data for SAC training neural networks is a 4-tuple $(\mathbf{s}_{l,k}, \mathbf{a}_{l,k}^{(1)}, r_{l,k}, \mathbf{s}_{l+1,k})$ stored in the replay buffer $\mathcal{M}$. During training, a batch size of $B$ is randomly selected from this buffer. The input layer of the AN receives the vector $\mathbf{s}_{l,k}$ and has two separate layers at the output, whose number of neurons is equal to the dimensions of the action. One layer represents the mean vector, and the other represents the standard deviation vector. An action vector with the following relation is generated using the obtained average and standard deviation vectors.	
\begin{equation}
[a^{(1)}_{l+1,k}]_{i} = [\mu_{\phib}(\mathbf{s}_{l,k})]_{i} + [\epsilon]_{i} [\sigma_{\phib}(\mathbf{s}_{l,k})]_{i}, \ i \in |\mathbf{a}_{l+1,k}^{(1)} |,
\end{equation}
where the sample vector $\epsilon$ is from  the common standard Gaussian distribution with dimensions equal to the action vector. The AN produces outputs $\mu_{\phib}$ and  $\sigma_{\phib}$ based on the input $\mathbf{s}_{l,k}$. Each action vector that is the output of an AN is obtained from a probability distribution. After receiving the action vector, the $\tanh$ function is applied to restrict the actions within the range of $-1$ to $1$. As mentioned before, the SAC algorithm is designed for an action space $\mathbf{a}_{l}^{(1)}$ with values $0$ and $1$, such that if $[a_{l+1,k}^{(1)}]_{i} > 0 $, $[a_{l+1,k}^{(1)}]_{i} = 1$, and if $[a_{l+1,k}^{(1)}]_{i} < 0$, then $[a_{l+1,k}^{(1)}]_{i} = -1$ for $i \in |\mathbf{a}_{l+1,k}^{(1)}|$.
CN$1$ and CN$2$ take the concatenated vector of state $\mathbf{s}_{l,k}$ and action $\mathbf{a}_{l,k}^{(1)}$ as input, and by passing them through the hidden layers, the output with a single neuron is produced that determines the value of $q_{\pi}(\mathbf{s}_{l,k}, \mathbf{a}_{l,k}^{(1)}; \thetab_{1})$ and $q_{\pi}(\mathbf{s}_{l,k}, \mathbf{a}_{l,k}^{(1)}; \thetab_{2})$. TCN$1$ and TCN$2$ take the connected vector of $\mathbf{s}_{l+1,k}$ and $\mathbf{a}_{l+1,k}^{(1)}$ (AN output) as input and offer $q_{\pi}(\mathbf{s}_{l+1,k}, \mathbf{a}_{l+1,k}^{(1)}; \bar{\thetab}_{1})$ and $q_{\pi}(\mathbf{s}_{l+1,k}, \mathbf{a}_{l+1,k}^{(1)}; \bar{\thetab}_{2})$ as output. 
}
\textcolor{black}{ The loss function, crucial for updating the parameters of the deep neural networks, is defined for updating CN1 parameters as follows
\begin{equation}\label{loss1}
\mathcal{L}(\thetab_{1}) = \frac{1}{\lvert B \rvert} \sum_{k=1}^{B} \big( y_{l+1,k} - q_{\pi} (\mathbf{s}_{l,k}, \mathbf{a}_{l,k}^{(1)}; \thetab_{1})  \big)^{2},
\end{equation}
where 
\begin{equation}
\begin{split}
y_{l+1,k} =& r_{l,k} + \gamma \Big[ \underset{i=1, 2}{\mathrm{min}} \quad q_{\pi}(\mathbf{s}_{l+1,k}, \mathbf{a}_{l+1,k}^{(1)}, \bar{\thetab}_{i})  \\
& -\lambda \log_{10} \Big( \pi \big( \mathbf{a}_{l+1,k} \big | \mathbf{s}_{l+1,k}; \phib \big) \Big)\Big].
\end{split}
\end{equation}
}

The loss function defined to update CN2 parameters is similar to the loss function applied to CN1 and is defined as follows:
\begin{equation}\label{loss2}
\mathcal{L}(\thetab_{2}) = \frac{1}{\lvert B \rvert} \sum_{k=1}^{B} \big( y_{l+1,k} - q_{\pi} (\mathbf{s}_{l,k}, \mathbf{a}_{l,k}^{(1)}; \thetab_{2})  \big)^{2}.
\end{equation}
The minimization of \eqref{loss1} and \eqref{loss2} can
be solved through application of the gradient descent technique, which involves updating $\thetab_{1}$ and $\thetab_{2}$ as 
\begin{equation}
\begin{split}
& \thetab_{1} = \thetab_{1} - \alpha_{1} \nabla_{\theta_{1}} \mathcal{L}(\thetab_{1}), \\
& \thetab_{2} = \thetab_{2} - \alpha_{2} \nabla_{\theta_{2}} \mathcal{L}(\thetab_{2}),
\end{split}
\end{equation}
where $0 < \alpha_{1} < 1$ and $0 < \alpha_{2} < 1$ represent the learning rates. The following loss function is employed for AN training.
\begin{equation}\label{parameters_SAC_updating}
\begin{split}
\mathcal{L}(\phib) =& \frac{1}{\lvert B \rvert} \sum_{k=1}^{B} \Big[ \underset{i=1, 2}{\mathrm{min}} \ q_{\pi} \Big(\mathbf{s}_{l,k}, \mu_{\phib}(\mathbf{s}_{l,k}) + \tilde{\epsilon}_{l,k} \sigma_{\phib}(\mathbf{s}_{l,k}), \thetab_{i}\Big) \\
& -\lambda \log_{10} \Big( \pi_{\phib}\big(\mu_{\phib}(\mathbf{s}_{l,k}) + \tilde{\epsilon}_{l,k} \sigma_{\phib}(\mathbf{s}_{l,k}) | \mathbf{s}_{l,k}; \phib\big) \Big)\Big].
\end{split}
\end{equation}
The parameters of AN are updated using the gradient descent method as 
\begin{equation}\label{parameter_of_AN_in_SAC}
\phib = \phib - \alpha_{3} \nabla_{\phi} \mathcal{L}(\phib)
\end{equation}
The parameters of TCN1 and TCN2 in the SAC-based algorithm are updated by
the soft update method, which are respectively expressed as
\begin{equation}\label{parameters_SAC_targets}
\begin{split}
& \bar{\thetab}_{1} = \tau \thetab_{1} + (1-\tau)\bar{\thetab}_{1}, \quad 0<\tau \leq 1,\\
& \bar{\thetab}_{2} = \tau \thetab_{2} + (1-\tau)\bar{\thetab}_{2}, \quad 0<\tau \leq 1.
\end{split}
\end{equation}

\begin{figure*}[ht] 
	\hspace*{-2.5cm} 	
	\centering
	\includegraphics[scale=0.50]{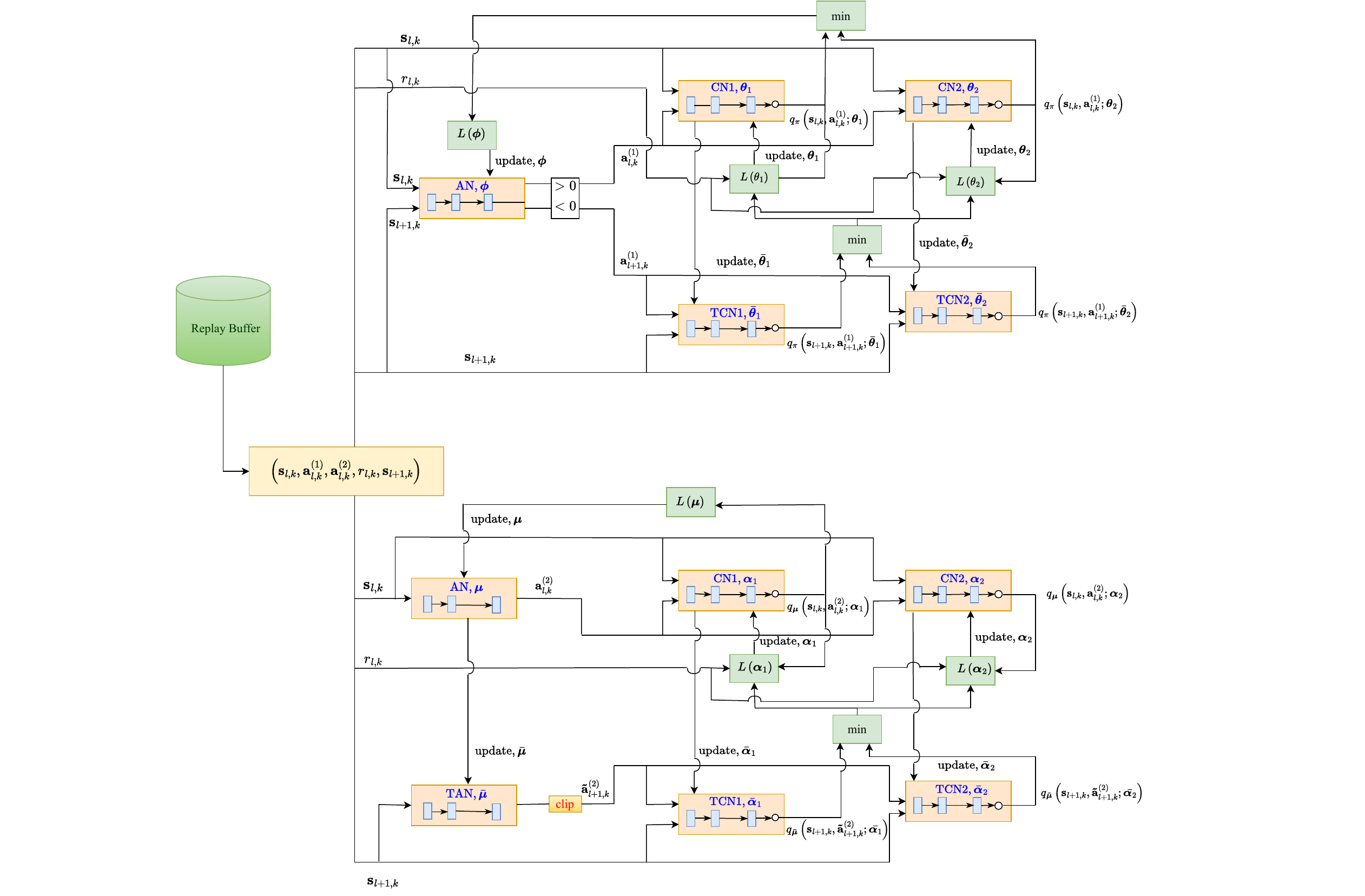}
	\caption{The proposed MSAT algorithm that simultaneously handles discrete and continuous action space. }
	\label{MSAT} 
\end{figure*}
\vspace*{-1em}
\section{TD$3$ Algorithm}
TD3 algorithm is a model-free, off-policy RL method, whose agent is an actor-critic RL agent that seeks an optimal policy aimed at maximizing the expected cumulative long-term reward \cite{TD3_algorithm}. The state-action value function is thus defined as follows
	\begin{equation}
	\begin{split}
	& q_{\mu}(\mathbf{s}_{l}, \mathbf{a}_{l}^{(2)}) =\\ & \mathbb{E}_{\text{Pr}(\mathbf{s}_{l+1}|\mathbf{s}_{l}, \mathbf{a}_{l}^{(2)})} \Big[ \sum_{l=0}^{\infty} \gamma ^{l} r(\mathbf{s}_{l}, \mu(\mathbf{s}_{l}) \big | \mathbf{s}_{l} = \mathbf{s}_{0}, \mathbf{a}_{l}^{(2)} = \mu(\mathbf{s}_{0}) \Big],
	\end{split}
	\end{equation}
	\normalsize
	where $\gamma \in (0, 1]$ denotes the discount factor. The optimal policy is given by
	\begin{equation}
	\mu^{\star}(\mathbf{s}_{l}) = \mathrm{arg} \ \underset{\mu(\mathbf{s}_{l}) \in \mathcal{A}}{\mathrm{max}} \quad q_{\mu}(\mathbf{s}_{l}, \mu(\mathbf{s}_{l})).
	\end{equation}
	The TD3 algorithm consists of six DNNs: the actor network (AN) with parameter $\mub$, critic network $1$ (CN$1$) with parameter $\alphab_{1}$, critic network $2$ (CN$2$) with parameter $\alphab_{2}$, target actor network $1$ (TAN) with parameter $\bar{\mub}$, target critic network $1$ (TCN$1$) with parameter $\bar{\alphab}_{1}$, and target critic network $2$ (TCN$2$) with parameter $\bar{\alphab}_{2}$. TD3 is an extended version of DDPG, which applies modifications to DDPG to control the overestimation of state-action value and prevent the sub-optimal policy generation. These modifications are detailed in the description of network training. 
	To train the DNNs within the structure of the TD3 algorithm, the training batch of training data $B$ $(\mathbf{s}_{l,k}, \mathbf{a}_{l,k}^{(2)}, r_{l,k}, \mathbf{s}_{l+1,k})$ is randomly selected from replay buffer $\mathcal{M}$. Upon inputting $\mathbf{s}_{l,k}$ into the AN, the output $\mathbf{a}_{l,k}^{(2)}$ is generated. CN1 and CN2 separately calculate $q_{\mub}(\mathbf{s}_{l,k}, \mathbf{a}_{l,k}^{(2)}; \alphab_{1})$ and $q_{\mub}(\mathbf{s}_{l,k}, \mathbf{a}_{l,k}^{(2)}; \alphab_{2})$  by receiving both $\mathbf{s}_{l,k}$ and $\mathbf{a}_{l,k}^{(2)}$. On the other hand, TAN generates output $\mathbf{a}_{l+1,k}^{(2)}$ after being given $\mathbf{s}_{l+1,k}$ from the replay buffer.
	The actions generated by TAN perturbs with noises for smoothing the corresponding state-action value function (first modification), thereby enhancing the residence of the policy against erroneous state-action function. The resulting smoothed target action is expressed as
	\begin{equation}
	\tilde{\mathbf{a}}_{l+1,k}^{(2)} = \text{clip}\big(\bar{\mub}(\mathbf{s}_{l+1,k}) + \text{clip}(\epsilon^{\prime}, -c, c), a_{\min}, a_{\max}  \big),
	\end{equation}	
where $\epsilon^{\prime} \sim (0, \sigma)$; and $c$ is the maximum exploration noise supported by the environment. TCN1 and TCN2 receive the state $\mathbf{s}_{l+1,k}$ and the action $\tilde{\mathbf{a}}_{l+1,k}^{(2)}$, producing outputs $q_{\bar{\mub}}(\mathbf{s}_{l+1,k}, \tilde{\mathbf{a}}_{l+1,k}^{(2)}; \bar{\alphab}_{1})$ and $q_{\bar{\mub}}(\mathbf{s}_{l+1,k}, \tilde{\mathbf{a}}_{l+1,k}^{(2)}; \bar{\alphab}_{2})$. The minimum of these outputs is utilized to learn the parameters $\alphab_{1}$ and $\alphab_{2}$. Incorporating two TCNs is considered the second modification on the structure of the TD3 compared to DDPG. The loss functions for updating the CN1 and CN2 parameters for $i \in \{1,2\}$ are given by
\begin{equation}\label{eq44}
	\mathcal{L}(\alphab_{i}) = \frac{1}{\lvert B \rvert} \sum_{k=1}^{B} \Big( q_{\mub}( \mathbf{s}_{l,k}, \mathbf{a}_{l,k}^{(2)}; \alphab_{i} ) - y(r_{l,k}, \mathbf{s}_{l+1,k})  \Big)^{2},
	\end{equation}
	where $	y(r_{l,k}, \mathbf{s}_{l+1,k}) = r_{l,k} + \gamma \underset{i=1,2}{\mathrm{min}} q_{\bar{\mub}} (\mathbf{s}_{l+1,k}, \tilde{\mathbf{a}}_{l+1,k}^{(2)}; \bar{\alphab}_{i}).$
To update CN1 and CN2 parameters, $\alphab_{1}, \alphab_{2}$, gradient descent algorithm is used on the loss function in \eqref{eq44} as 
\begin{equation}\label{parameters_of_TD3_critics}
\alphab_{i} = \alphab_{i} - \theta_{i} \nabla_{\alpha_{i}} \mathcal{L}(\alphab_{i}), \ i=1,2,
\end{equation}
where $0 < \theta_{i} < 1$ denotes learning rate.
The last modification in the TD3 agent is updating the AN and TAN parameters less frequently than the other networks. In the calculation of loss function AN, only $q_{\mu}(\mathbf{s}_{l,k}, \mathbf{a}_{l,k}^{(2)}, \alphab_{1})$ is used and it is defined as follows
\begin{equation}
\mathcal{L}(\mub) = \frac{1}{\lvert B \rvert} \sum_{k=1}^{B} q_{\mu}( \mathbf{s}_{l,k}, \mathbf{a}_{l,k}^{(2)}, \alphab_{1} ).
\end{equation}
Gradient descent can also be used to update AN parameters.
\begin{equation}\label{parameter_TD3_AN}
\mub = \mub - \theta_{3} \nabla_{\mu} \mathcal{L}(\mub).
\end{equation}
The TAN, TCN1, and TCN2 are refreshed
based on the AN, CN1, and CN2 
\begin{equation}\label{parameters_TD3_targets}
\begin{split}
& \bar{\alphab}_{i} = \tau_{\alpha} \alphab_{i} + (1 - \tau_{\alpha}) \bar{\alphab}_{i}, \ \ i = 1, 2,\\
& \bar{\mub} = \tau_{\mu} \mub + (1 - \tau_{\mu}) \bar{\mub},
\end{split}
\end{equation}
where $\tau_{\alpha} $ and $\tau_{\mu}$ are the decaying rates for the AN and CNs, respectively. The pseudo-code for the proposed MSAT algorithm is outlined in Algorithm $1$. 

\begin{algorithm}[h!]
	\caption{: The Proposed MSAT Algorithm}
	\begin{algorithmic}[1]
		\State \textbf{Input:} 
		Number of episodes $E$, number of time slots $L$, and batch size $B$.
		\normalsize \State \textbf{Initialization:} Randomly initialize $\phib$, $\thetab_{1}$, and $\thetab_{2}$ for the modified-SAC algorithm and $\mub$, $\alphab_{1}$, and $\alphab_{2}$ for the TD$3$ algorithm. Initialize the SAC TCNs with $\bar{\thetab}_{1} \leftarrow \thetab_{1}$ and $\bar{\thetab}_{2} \leftarrow \thetab_{2}$, as well as TAN and TCNs for TD$3$ algorithm with $\bar{\mub} \leftarrow \mub$, $\bar{\alphab}_{1} \leftarrow \alphab_{1}$, and $\bar{\alphab}_{2} \leftarrow \alphab_{2}$.
		\For{\texttt{$e = 1, 2, \dots, E$}}
		\State Reset environment to obtain the initial state.
		\For{\texttt{$l = 1, 2, \dots, L$}}
		\State Observe state $\mathbf{s}_{l}$ and select action $\mathbf{a}_{l}$.
		\State Execute action $\mathbf{a}_{l}$ on the environment.
		\State Observe next state $\mathbf{s}_{l+1}$ and  receive reward $r_{l}$.
		\State Store $\left(\mathbf{s}_l, \mathbf{a}_l, r_l, \mathbf{s}_{l+1} \right)$ in $\mathcal{M}$.
		\State {Randomly sample a batch set $B$ from $\mathcal{M}$.}
		\State \small Update the network parameters $\phib$, $\thetab_{1}$, $\thetab_{2}$, $\mub$, $\alphab_{1}$, and $\alphab_{2}$ using \eqref{parameters_SAC_updating}, \eqref{parameter_of_AN_in_SAC}, \eqref{parameters_of_TD3_critics}, and \eqref{parameter_TD3_AN}.
		\small \State Update the target network parameters $\bar{\thetab}_{1}$, $\bar{\thetab}_{2}$, $\bar{\mub}$, $\bar{\alphab}_{1}$, and $\bar{\alphab}_{2}$ using \eqref{parameters_SAC_targets} and \eqref{parameters_TD3_targets}.
		\EndFor
		\EndFor
	\end{algorithmic}
\end{algorithm}

\begin{figure*}[ht] 
	\hspace*{0.6cm} 	
	\centering
	\includegraphics[scale=0.41]{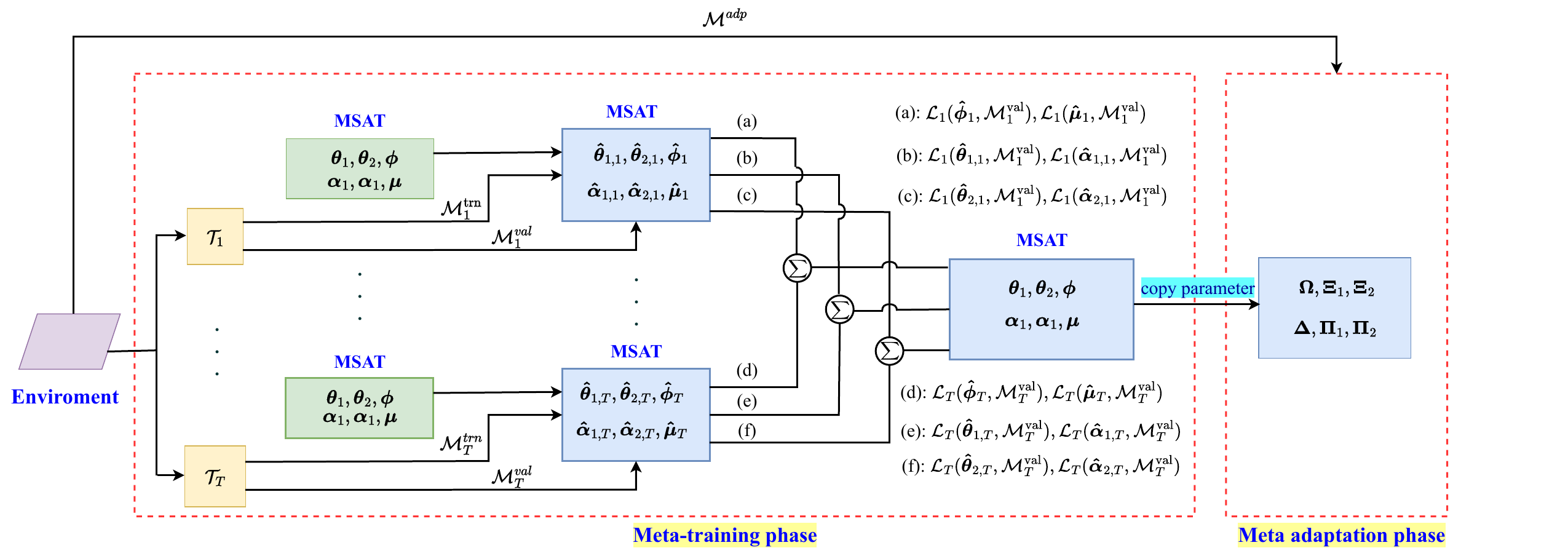}
	\caption{Overall schematic of proposed MMSAT algorithm.}
	\label{MMSAT} 
\end{figure*}

\vspace{-1 em}
\section{Proposed Meta-Modified-SAC-TD3 (MMSAT) Algorithm}
\textcolor{black}{
The problem addressed in this paper involves a dynamic environment of users and AARIS. As stated earlier, the optimization problem $\mathcal{P}_{0}$ is non-convex. The traditional DRL methods including SAC and TD3 can handle dynamic settings, but they assume no changes between training and testing environments, which is impractical for scenarios with mobile users/AARIS. Recently, it has been demonstrated that meta-learning can effectively address such mismatch issues in wireless networks~\cite{Meta1, Meta_Faramarzi, Meta2}.
To take advantage of meta-learning, specifically the MAML approach proposed in \cite{MAML}, we integrate meta-learning with modified-SAC and TD3, introducing the MMSAT algorithm for solving problem $\mathcal{P}_{0}$.
In the MSAT algorithm, only one task is considered, while
the main idea of MMSAT is to grant the learning model
adaptability to new tasks by considering the MDP problem as a
meta-task set $\mathcal{T}$ with $T$ different tasks and their experience memory $\mathcal{M}_{t}$, $t = 1, . . . , T$. Hence, problem $\mathcal{P}_{0}$ can be assumed as a meta-task set, where each task is defined by considering different initial locations for users. To clarify, the number of meta-tasks in problem $\mathcal{P}_{0}$ is equivalent to the number of times that users are randomly being placed in an initial location.
}

\textcolor{black}{
The proposed MMSAT algorithm consists of two main
phases, namely \textit{meta-training} and \textit{meta-adaptation}. In the meta-training phase, training and evaluation of the learning
model are performed. Particularly, for the model training, a batch set denoted by $\mathcal{M}_{t}^{\mathrm{trn}}$ is randomly sampled from  $\mathcal{M}_{t}$. Additionally,  $\mathcal{M}_{t}^{\mathrm{trn}}$ is used to update the parameters of the actor and critic networks for modified-SAC and TD3 for each task $ t $ using the stochastic gradient descent (SGD) algorithm \cite{SGD}. Moreover, another randomly sampled batch set $\mathcal{M}_{t}^{\mathrm{val}}$ is used to update the parameters of the global model over all tasks $T$.} 

In the meta-adaptation phase, the performance of MMSAT is evaluated by introducing a new agent equipped with an actor network, two critic networks, and two target critic networks for the modified-SAC, and 
an actor network, two critic networks, a TAN, and two target critic networks for the TD3. To train the new agent, a  new task is defined with users placed randomly. The difference between training the new agent in the meta-adaptation phase and the agent in the meta-training phase lies in the fact that the parameters of the  model in the meta-adaptation phase are initialized with the obtained parameters from the meta-training phase rather than random initialization.  
Furthermore, meta-adaptation involves utilizing an experience memory  $\mathcal{M}_{\mathrm{adp}}$ to store the experienced transitions, including the current state, next state, chosen action, and received reward of the new agent. The pseudo-code for the proposed MMSAT algorithm is outlined in Algorithm 2.	

In the meta-training phase, for each task $ t $, the environment is reset to obtain the initial state $ \mathbf{s}^t_1 $. Then, at each time slot $ l $, the agent selects an action $ \mathbf{a}^t_l $ and executes it on the environment. As a result, the agent receives reward $ r^t_l $ and transits to the next state $ \mathbf{s}^t_{l+1} $. The experienced transition $ \left(\mathbf{s}^t_l, \mathbf{s}^t_{l+1}, \mathbf{a}^t_l, r^t_l  \right) $ is stored in the experience memory $ \mathcal{M}_t $.  When the number of experienced transitions stored in $ \mathcal{M}_t $ reaches or exceeds the batch size, a  batch set of $ \mathcal{M}_t^{\mathrm{trn}} $  is randomly sampled from $ \mathcal{M}_t$. 
Exploiting this sampled batch $ \mathcal{M}_t^{\mathrm{trn}} $ , the parameters of the actor and critic networks in modified-SAC and TD3 for each task $ t $ are updated, denoted by $\hat{\phib}_{t}$,  $\hat{\thetab}_{1, t}$, $\hat{\thetab}_{2, t}$, $\hat{\mub}_{t}$,  $\hat{\alphab}_{1, t}$, and $\hat{\alphab}_{2, t}$. The loss functions for training these DNNs are defined as
\vspace*{0.1em}
\begin{align}\label{task_parameter_SAC}
&\left\lbrace \begin{array}{lc}
\hat{\phib}_{t} = \mathrm{arg} \ \underset{\phi}{\mathrm{min}} \quad \mathcal{L}_{t}(\mathbf{\phib}, \mathcal{M}_t^{\mathrm{trn}} ),\\
\hat{\thetab}_{1,t} = \mathrm{arg} \ \underset{\theta_{1}}{\mathrm{min}} \quad \mathcal{L}_{t}(\mathbf{\thetab_{1}}, \mathcal{M}_t^{\mathrm{trn}} ) ,\\
\hat{\thetab}_{2,t} = \mathrm{arg} \ \underset{\theta_{2}}{\mathrm{min}} \quad \mathcal{L}_{t}(\mathbf{\thetab_{2}}, \mathcal{M}_t^{\mathrm{trn}} ) .
\end{array}\right. 
\end{align}
\normalsize
and
\begin{align}\label{task_parameter_TD3}
&\left\lbrace \begin{array}{lc}
\hat{\mub}_{t} = \mathrm{arg} \ \underset{\mu}{\mathrm{min}} \quad \mathcal{L}_{t}(\mathbf{\mub}, \mathcal{M}_t^{\mathrm{trn}} ),\\
\hat{\alphab}_{1,t} = \mathrm{arg} \ \underset{\alpha_{1}}{\mathrm{min}} \quad \mathcal{L}_{t}(\mathbf{\alphab_{1}}, \mathcal{M}_t^{\mathrm{trn}} ) ,\\
\hat{\alphab}_{2,j} = \mathrm{arg} \ \underset{\alpha_{2}}{\mathrm{min}} \quad \mathcal{L}_{t}(\mathbf{\alphab_{2}}, \mathcal{M}_t^{\mathrm{trn}} ) .
\end{array}\right. 
\end{align}
\normalsize
After accomplishing the aforementioned steps for all tasks $ t\in\mathcal{T} $, the evaluation step is performed. Specifically, we sample a batch set of  $\mathcal{M}_{t}^{\mathrm{val}}$ from $\mathcal{M}_{t}$. The global parameters of the actor network ($\phib$) and critic networks ($\thetab_{1}$ and $\thetab_{2}$) for modified-SAC, as well as  
AN ($\mub$), CN1 and CN2 ($\alphab_{1}$ and $\alphab_{2}$) for TD3
are calculated using the derivative of the corresponding loss functions $\sum_{t} \mathcal{L}_{t} (\hat{\phib}_{t}, \mathcal{M}_{t}^{\mathrm{val}})$, $\sum_{t} \mathcal{L}_{t} (\hat{\thetab}_{1, t}, \mathcal{M}_{t}^{\mathrm{val}}) $, $\sum_{t} \mathcal{L}_{t} (\hat{\thetab}_{2, t}, \mathcal{M}_{t}^{\mathrm{val}}) $, 
$\sum_{t} \mathcal{L}_{t} (\hat{\mub}_{t}, \mathcal{M}_{t}^{\mathrm{val}})$, $\sum_{t} \mathcal{L}_{t} (\hat{\alphab}_{1, t}, \mathcal{M}_{t}^{\mathrm{val}}) $, and $\sum_{t} \mathcal{L}_{t} (\hat{\alphab}_{2, t}, \mathcal{M}_{t}^{\mathrm{val}}) $. Thus, the optimization problems
used to optimize $\mathbf{\phib}$, $\mathbf{\thetab}_{1}$, $\mathbf{\thetab}_{2}$, $\mathbf{\mub}$, $\mathbf{\alphab}_{1}$, and $\mathbf{\alphab}_{2}$ can be respectively expressed as
\begin{align}\label{global_SAC}
&\left\lbrace \begin{array}{lc}
\mathbf{\phib} = \mathrm{arg} \ \underset{\phi}{\mathrm{min}} \quad \sum_{t} \mathcal{L}_{t} (\hat{\phib}_{t}, \mathcal{M}_{t}^{\mathrm{val}}),\\
\thetab_{1} = \mathrm{arg} \ \underset{\theta_{1}}{\mathrm{min}} \quad \sum_{t} \mathcal{L}_{t} (\hat{\thetab}_{1, t}, \mathcal{M}_{t}^{\mathrm{val}}) ,\\
\thetab_{2} = \mathrm{arg} \ \underset{\theta_{2}}{\mathrm{min}} \quad \sum_{t} \mathcal{L}_{t} (\hat{\thetab}_{2, t}, \mathcal{M}_{t}^{\mathrm{val}}).
\end{array}\right. 
\end{align}
and
\begin{align}\label{global_TD3}
&\left\lbrace \begin{array}{lc}
\mathbf{\mub} = \mathrm{arg} \ \underset{\mu}{\mathrm{min}} \quad \sum_{t} \mathcal{L}_{t} (\hat{\mub}_{t}, \mathcal{M}_{t}^{\mathrm{val}}),\\
\alphab_{1} = \mathrm{arg} \ \underset{\alpha_{1}}{\mathrm{min}} \quad \sum_{t} \mathcal{L}_{t} (\hat{\alphab}_{1, t}, \mathcal{M}_{t}^{\mathrm{val}}) ,\\
\alphab_{2} = \mathrm{arg} \ \underset{\alpha_{2}}{\mathrm{min}} \quad \sum_{t} \mathcal{L}_{t} (\hat{\alphab}_{2, t}, \mathcal{M}_{t}^{\mathrm{val}}).
\end{array}\right. 
\end{align}
\vspace{-0.1em}
Using the meta-training phase, we have trained a learning model on all meta-tasks. The parameters of MSAT algorithm are used for training another learning model on a new task in the meta-adaptation phase.  
Similar to the meta-training phase, the meta-adaptation phase uses an actor and two critic networks with parameters of $\mathbf{\Omega}$, $\mathbf{\Xi}_{1}$, and $\mathbf{\Xi}_{1}$ for modified-SAC and with parameters of $\mathbf{\Delta}$, $\mathbf{\Pi_{1}}$, and $\mathbf{\Pi_{2}}$ for TD$3$, respectively. At the initial step, the network parameters are replaced by the corresponding parameters of the learning model in the training phase. That is, $\mathbf{\Omega} \leftarrow \phib$, $\mathbf{\Xi}_{1} \leftarrow \thetab_{1}$, and $\mathbf{\Xi}_{2} \leftarrow \thetab_{2}$ for modified-SAC algorithm and 
$ \mathbf{\Delta} \leftarrow \mub$, $\mathbf{\Pi}_{1} \leftarrow \alphab_{1}$, and $\mathbf{\Pi}_{2} \leftarrow \alphab_{2}$ for TD$3$ algorithm.
For meta-adaptation phase, there is another replay buffer as $\mathcal{M}^{\mathrm{ada}}$ of
which the agent randomly samples a batch of it and updates the parameters of actor and critic networks for modified-SAC and TD$3$ as
\begin{align}\label{eq21}
\left\{ \begin{array}{l}
\mathbf{\Omega} = \mathbf{\Omega}-\alpha_{1}{\nabla _{\Omega} }{{\mathcal{L}_{s}}({\mathbf{\Omega}}, \mathcal{M}^{\mathrm{ada}})},\\
\mathbf{\Xi}_{1} = \mathbf{\Xi}_{1}-\alpha_{2}{\nabla _{\Xi_{1}} }{{\mathcal{L}_{s}}({\mathbf{\Xi}_{1}}, \mathcal{M}^{\mathrm{ada}})},\\
\mathbf{\Xi}_{2} = \mathbf{\Xi}_{2}-\alpha_{3}{\nabla _{\Xi_{2}} }{{\mathcal{L}_{s}}({\mathbf{\Xi}_{2}}, \mathcal{M}^{\mathrm{ada}})},
\end{array} \right.
\end{align}
where $\alpha_{1}$, $\alpha_{2}$, and $\alpha_{3}$ represent the learning rate of the
modified-SAC.
\begin{align}\label{eq55}
\left\{ \begin{array}{l}
\mathbf{\Delta} = \mathbf{\Delta}-\theta_{1}{\nabla _{\Delta} }{{\mathcal{L}_{t}}({\mathbf{\Delta}}, \mathcal{M}^{\mathrm{ada}})},\\
\mathbf{\Pi}_{1} = \mathbf{\Pi}_{1}-\theta_{2}{\nabla _{\Pi_{1}} }{{\mathcal{L}_{t}}({\mathbf{\Pi}_{1}}, \mathcal{M}^{\mathrm{ada}})},\\
\mathbf{\Pi}_{2} = \mathbf{\Pi}_{2}-\theta_{3}{\nabla _{\Pi_{2}} }{{\mathcal{L}_{t}}({\mathbf{\Pi}_{2}}, \mathcal{M}^{\mathrm{ada}})},
\end{array} \right.
\end{align}
where $\theta_{1}$, $\theta_{2}$, and $\theta_{3}$ represent the learning rate of the
TD$3$.
Using the parameters resulting from the meta-training phase leads to a better  generalization ability of the MMSAT compared to the MSAT algorithm. The architecture of the MSAT algorithm is shown in Fig. \ref{MMSAT}.

\begin{algorithm}[h!]
	\caption{: The Proposed MMSAT Algorithm}
	\begin{algorithmic}[1]
		\State \textbf{Input:} 
		Number of episodes in meta-raining and meta-adaptation phases  $E_{\mathrm{trn}}$ and $E_{\mathrm{adp}}$, number of time slots $L$, and number of tasks $T$.
		\small \State \textbf{Initialization:}
		Randomly initialize $\phib$, $\thetab_{1}$, and $\thetab_{2}$ for the modified-SAC algorithm as well as $\mub$, $\alphab_{1}$, and $\alphab_{2}$ for the TD$3$ algorithm. Initialize the SAC TCNs with $\bar{\thetab}_{1} \leftarrow \thetab_{1}$ and $\bar{\thetab}_{2} \leftarrow \thetab_{2}$, and TAN and TCNs for TD$3$ algorithm with $\bar{\mub} \leftarrow \mub$, $\bar{\alphab}_{1} \leftarrow \alphab_{1}$, and $\bar{\alphab}_{2} \leftarrow \alphab_{2}$. Initializing empty experience memory $\mathcal{M}_{t}$ for each task $ t \in\mathcal{T} $, and empty experience memory $\mathcal{M}_{\mathrm{adp}}$. 
		\normalsize \algrule
		\begin{center}
			\textbf{Meta-training phase}
		\end{center}
		\For{\texttt{$e = 1, 2, \dots, E_{\mathrm{trn}}$}}
		\State Reset environment to obtain the initial state.
		\For{\texttt{$l = 1, 2, \dots, L$}}
		\For{\texttt{$t = 1, 2, \dots, T$}}
		\State Observe state $\mathbf{s}_{l}^{t}$ and select action $\mathbf{a}_{l}^{t}$.
		\State Execute action $\mathbf{a}_{l}$ on the environment.
		\State Observe next state $\mathbf{s}_{l+1}^{t}$ and  receive reward $r_{l}^{t}$.
		\State Store $\left(\mathbf{s}_l^{t}, \mathbf{a}_l^{t}, r_l^{t}, \mathbf{s}_{l+1}^{t} \right)$ in $\mathcal{M}_{t}$.
		\State {Randomly sample a batch set $\mathcal{M}_{t}^{\mathrm{trn}}$ from $\mathcal{M}_{t}$.}
		\State Update network parameters $\hat{\phib}_{t}$, $\hat{\thetab}_{1, t}$, and $\hat{\thetab}_{2, t}$ using gradient descent for modified-SAC algorithm by \eqref{task_parameter_SAC}.
		\State Update network parameters $\hat{\mub}_{t}$, $\hat{\alphab}_{1, t}$, and $\hat{\alphab}_{2, t}$ using gradient descent for TD$3$ algorithm by \eqref{task_parameter_TD3}.
		\EndFor
		\State {Randomly sample a batch set $\mathcal{M}_{t}^{\mathrm{val}}$ from $\mathcal{M}_{t}$.}
		\small \State Calculate  the gradient of loss functions $\sum_{t} \mathcal{L}_{t} (\hat{\phib}_{t}, \mathcal{M}_{t}^{\mathrm{val}})$, $\sum_{t} \mathcal{L}_{t} (\hat{\thetab}_{1, t}, \mathcal{M}_{t}^{\mathrm{val}}) $, $\sum_{t} \mathcal{L}_{t} (\hat{\thetab}_{2, t}, \mathcal{M}_{t}^{\mathrm{val}}) $, $\sum_{t} \mathcal{L}_{t} (\hat{\mub}_{t}, \mathcal{M}_{t}^{\mathrm{val}})$, $\sum_{t} \mathcal{L}_{t} (\hat{\alphab}_{1, t}, \mathcal{M}_{t}^{\mathrm{val}}) $, and $\sum_{t} \mathcal{L}_{t} (\hat{\alphab}_{2, t}, \mathcal{M}_{t}^{\mathrm{val}}) $ for updating $\phib$, $\thetab_{1}$, $\thetab_{2}$, $\mub$, $\alphab_{1}$, and $\alphab_{2}$ in $\eqref{global_SAC}$ and $\eqref{global_TD3}$ using gradient descent.
		\normalsize \EndFor
		\State Update the global model parameters $\phib$, $\thetab_{1}$, $\thetab_{2}$, $\mub$, $\alphab_{1}$, and $\alphab_{2}$ using gradient descent over $\eqref{global_SAC}$ and $\eqref{global_TD3}$.
		\EndFor
		\algrule
		\begin{center}
			\textbf{Meta-adaptation phase}
		\end{center}
		\State Initialize $\mathbf{\Omega} \leftarrow \phib$, $\mathbf{\Xi}_{1} \leftarrow \thetab_{1}$, and $\mathbf{\Xi}_{2} \leftarrow \thetab_{2}$ for modified-SAC algorithm.
    	\State Initialize $ \mathbf{\Delta} \leftarrow \mub$, $\mathbf{\Pi}_{1} \leftarrow \alphab_{1}$, and $\mathbf{\Pi}_{2} \leftarrow \alphab_{2}$ for TD$3$ algorithm.
		\For{$e = 1, 2, \dots, E_{\mathrm{adp}}$}
		\State Reset environment to obtain the initial state.
		\For{$l = 1, 2, \dots, L$}			
		\State Generate  experienced transitions $\left(\mathbf{s}_l, \mathbf{a}_l, r_l, \mathbf{s}_{l+1} \right)$ similar to steps $7$ to $9$ in the meta-training phase.
		\State Store the experienced transitions in $\mathcal{M}^{\mathrm{ada}}$ and randomly sample a batch set.
		\State Update the network parameters of the modified-SAC algorithm $\mathbf{\Omega}$, $\mathbf{\Xi}_{1}$, and $\mathbf{\Xi}_{2}$.
		\State Update the network parameters of the TD$3$ algorithm $\mathbf{\Delta}$, $\mathbf{\Pi}_{1}$, and $\mathbf{\Pi}_{2}$.
		\EndFor
		\EndFor
	\end{algorithmic}
\end{algorithm}
\vspace{-1 em}
\begin{figure*}[h!]
	\centering
	\begin{subfigure}[b]{0.3\textwidth}
		\centering
		\includegraphics[width=\textwidth]{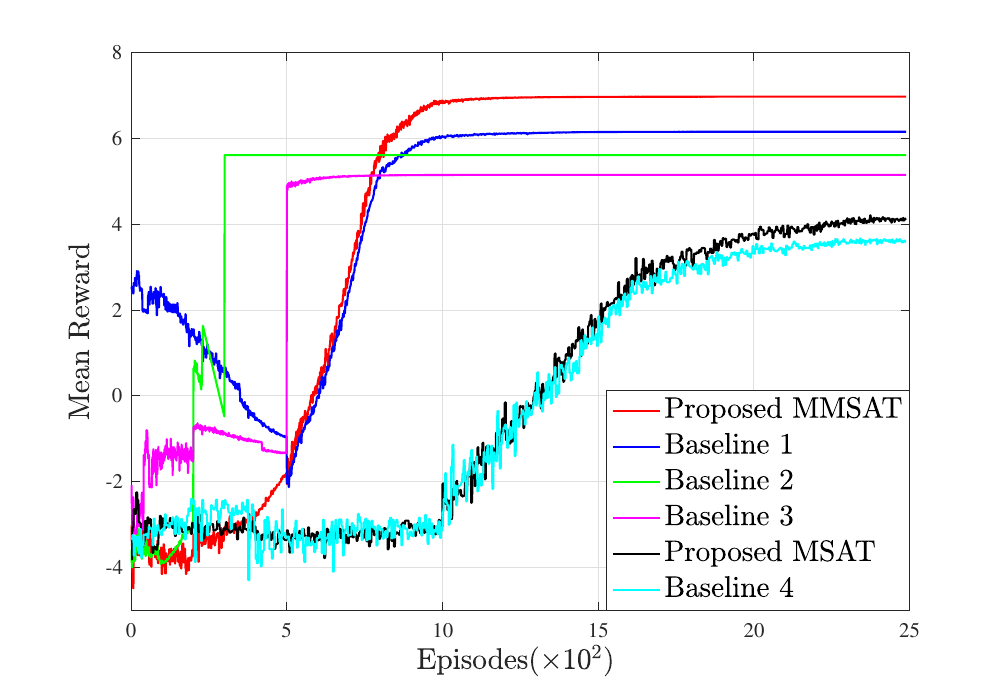}
		\caption{Mean reward.}
		\label{reward_episode} 
	\end{subfigure}
	\hspace*{0.25cm}
	\begin{subfigure}[b]{0.3\textwidth}
		\centering
		\includegraphics[width=\textwidth]{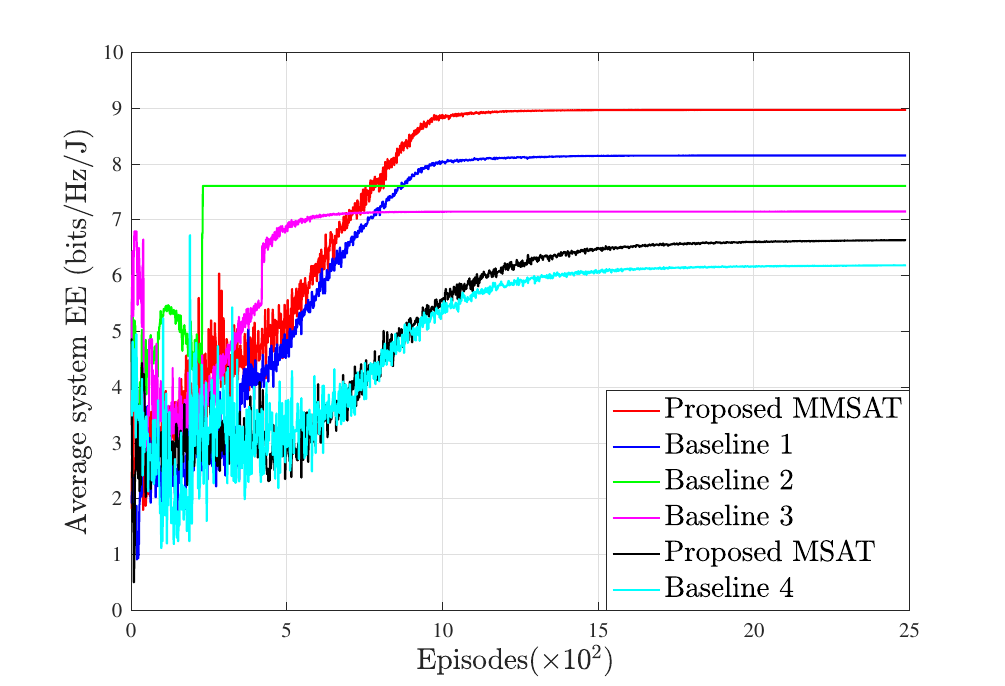}
		\caption{Average EE.}
		\label{EE_episode}
	\end{subfigure}
	\begin{subfigure}[b]{0.3\textwidth}
		\centering
		\includegraphics[width=\textwidth]{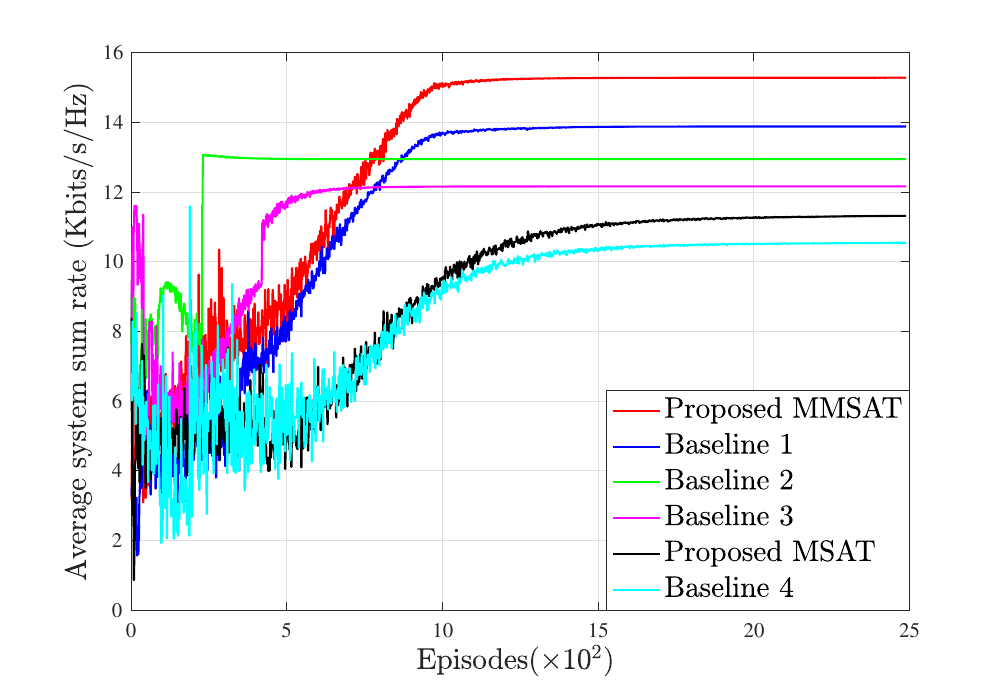}
		\caption{Average sum rate.}
		\label{sumrate_episode} 
	\end{subfigure}
	\caption{Convergence behaviour of the proposed MMSAT and baselines.}
	\label{convergence}
\end{figure*}
\vspace*{-0.5em}
\section{Complexity Analysis}
In this section, we present the computational complexity of the proposed MMSAT resource management algorithm. The computational complexity of the MMSAT is in terms of both meta-training and meta-adaptation phases. The former, i.e., the meta-training is of computational complexity of:
\begin{equation}
\mathcal{O} \Big( \big( \sum_{l=0}^{N_{L}} h_{l} h_{l+1} \big) \times \mathcal{M}_{t}^{\mathrm{trn}}  \times E_{\mathrm{trn}} \times L \times T\Big),
\end{equation}
where $N_{L}$, $h$, and $h_{l}$ denote the number of neural network layers, number of neurons in each
layer, and number of neurons in the $l$-th layer. As well, the number of time slots and tasks are represented by $L$ and $T$, respectively. Next, we calculate the computational complexity of the latter, i.e., the meta-adaptation phase
as follows. 
\begin{equation}
\mathcal{O} \Big( \big( \sum_{l=0}^{N_{L}} h_{l} h_{l+1} \big) \times \mathcal{M}_{t}^{\mathrm{adp}}  \times E_{\mathrm{adp}} \times L\Big).
\end{equation}
The overall computational complexity of the MMSAT algorithm mainly depends on the latter phase.
\section{Simulation Results}
In this section, the performance of the considered energy efficient AARIS-aided system with RSMA is evaluated  via numerical analysis. 
\vspace{-1 em}
\subsection{System Configuration}
The network design comprises of one AARIS, one BS, and eight single antenna users. More specifically, a single active RIS with $M = \lbrace9,16,25,36\rbrace$ number of active reflective elements is mounted on a UAV.  Moreover, a three 3D coordinate system is considered, where a BS is located at $(0,0,10)$ m, while the minimum and maximum locations of AARIS are respectively considered as $\mathbf{q}_\textrm{min}=(0,0,100)$ m and $\mathbf{q}_\textrm{max}=(150,150,100)$ m. The BS with $N_{\text{BS}} = \lbrace 3,5,7,11\rbrace$ number of antennas serves users which are uniformly distributed between a rectangular area with $(0,0)$ and $(150, 150)$ cordinates. 
The simulation parameters are summarized in Table \ref{tab:my_label}. 
\vspace{-1 em}
\subsection{Convergence Behaviour}
Figs. \ref{reward_episode}-\ref{sumrate_episode} depict the convergence behaviour of the baselines within $2500$ episodes. To evaluate the performance of the AARIS system, the mean reward, EE, as well as the sum rate of the AARIS system are adopted as the metrics. We consider the four benchmark schemes for performance comparison, which are:
\begin{enumerate}
	\item \textbf{Proposed MMSAT}: This baseline corresponds to the \textbf{Algorithm 2} and adopts meta-learning, modified-SAC and TD3 with RSMA.
	\item \textbf{Proposed MSAT}: This baseline corresponds to the \textbf{Algorithm 1} and adopts modified-SAC and TD3 with RSMA.
	\item \textbf{Baseline 1}: This baseline adopts meta-learning, modified-SAC and TD3 with NOMA.
	\item \textbf{Baseline 2}: This baseline adopts meta-learning, modified-SAC and TD3 with RSMA, however with passive RIS.
	\item \textbf{Baseline 3}: This baseline adopts meta-learning, modified-SAC and TD3 with RSMA, however with a fixed terrestrial active RIS at $(75, 75, 100)$ m.
	\item \textbf{Baseline 4}: This baseline is similar to the \textbf{proposed MSAT} baseline, but the former adapts NOMA to serve the users.
\end{enumerate}
One can capture the following observations from the abovementioned figures. 
At a glance, it is perceived that the convergence for the baselines including the meta-learning is yielded much faster, due to better adaptability and also more observations from the system dynamics. 
 The \textbf{proposed MSAT} and \textbf{Baseline 4}, which lack the meta-learning, exhibit worse performance compared to others including meta-learning. The convergence speed for the \textbf{Baseline 3} is higher than that for rest of the baselines, due to lacking the optimization of trajectory and velocity of the UAV, and thereby, smaller action space $\mathbf{a}_{l}^{(2)}$ for the DRL. 
 Figs. \ref{reward_episode}-\ref{sumrate_episode} can also be analyzed from the convergence speed perspective. Namely, the smaller the action space of the DRL is, the faster convergence is achieved. Amongst others, the \textbf{Baseline 2} benefits from faster convergence in comparison with \textbf{proposed MMSAT} and \textbf{Baseline 1}, thanks to lacking the optimization of the RIS amplification factor matrix, and thereupon, smaller DRL action space $\mathbf{a}_{l}^{(2)}$.
 Another observation we found is that the baselines including RSMA, which we proposed for the AARIS system in this paper, outperform the ones including NOMA, which was proposed for an ARIS system in \cite{NOMA1,NOMA2}, thanks to more efficient interference management of the RSMA. 
 Figs. \ref{reward_episode}-\ref{sumrate_episode} also unveil that our AARIS system in \textbf{proposed MMSAT} baseline with active RIS outperforms the ARIS system corresponded to the \textbf{Baseline 2} in prior literature \cite{Surv1,Surv2} with passive RIS, for all evaluation metrics. This is due to the fact that incorporating an active RIS achieves more SE, compared to the passive type. Similarly, the baseline corresponded to our proposed AARIS system is superior to the fixed terrestrial RIS, where the RIS is not mounted at the UAV. This understanding validates similar observations in \cite{Surv1} and the references therein. 
 Eventually, by comparing the \textbf{Baseline 2} and \textbf{Baseline 3}, it can be concluded that an aerial RIS-assisted communication with a passive RIS is more efficient than a fixed terrestrial RIS-assisted communication even with an active RIS, as shown for all evaluation metrics.

\begin{figure*}[h!]
	\centering
	\begin{subfigure}[b]{0.3\textwidth}
	\centering
    \includegraphics[width=\textwidth]{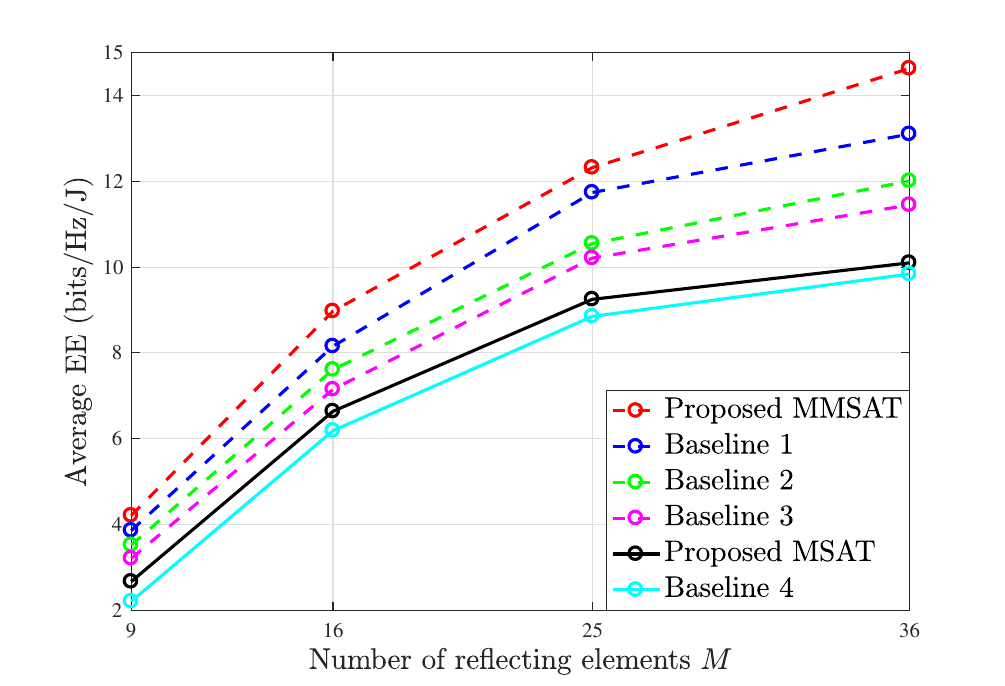}
    \caption{\small Average EE vs. the number of
	RIS elements $M$.}
    \label{EE_elements} 
	\end{subfigure}
	\hspace*{0.25cm}
	\begin{subfigure}[b]{0.3\textwidth}
	\centering
    \includegraphics[width=\textwidth]{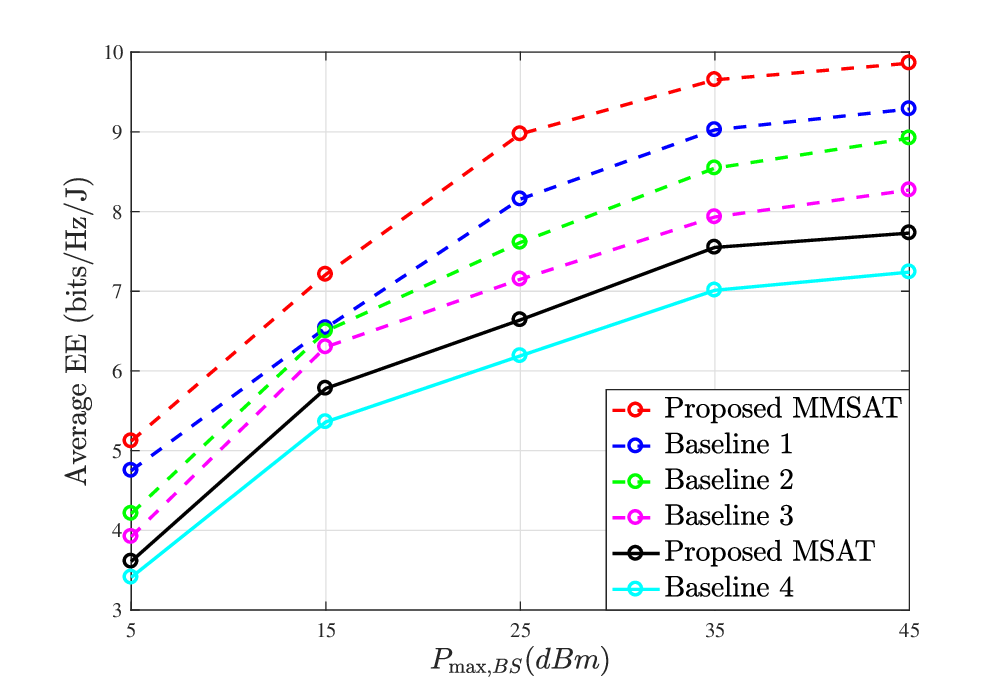}
    \caption{\small Average EE vs. the BS transmit power budget $P_{\text{max},\text{BS}}$.}
    \label{EE_power} 
	\end{subfigure}
    \hspace*{0.25cm}
    \begin{subfigure}[b]{0.3\textwidth}
    \centering
    \includegraphics[width=\textwidth]{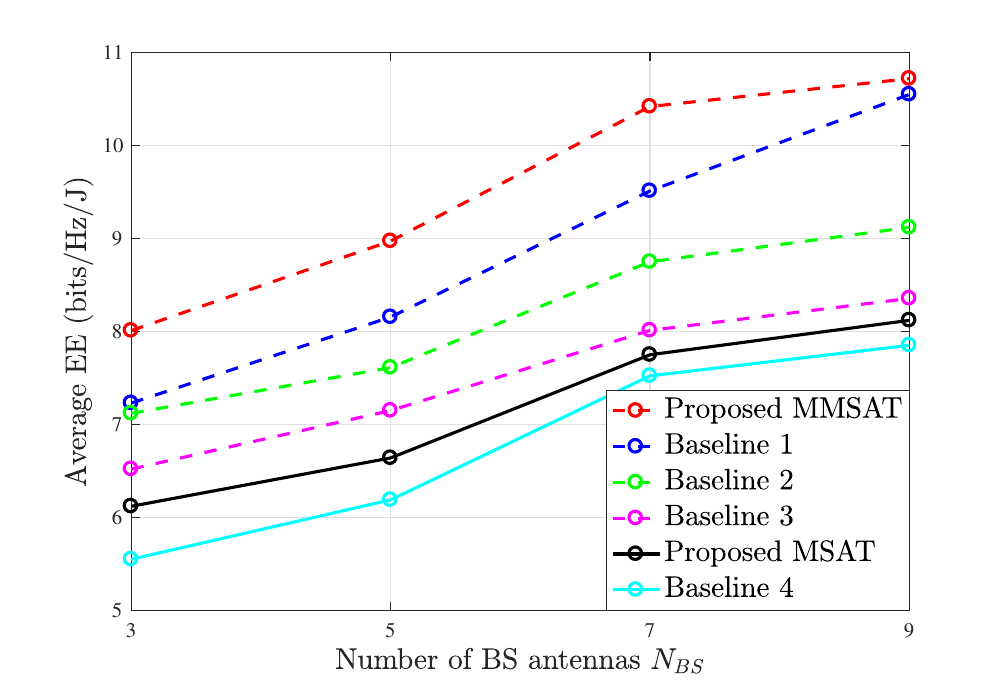}
    \caption{\small Average EE vs. the number of
    BS antennas $N_\text{BS}$.}
    \label{N_BS} 
    \end{subfigure}
	\caption{\small Impact of system parameters on overall average system EE.}
	\label{convergence}
\end{figure*}

\vspace{-1 em}
\subsection{Impact of System Parameters}
Figs. \ref{EE_elements}-\ref{N_BS} respectively portray the impact of number of RIS elements, as well as the transmit power budget and the number of antennas at the BS, on average system EE, by letting $N_{\text{BS}}~=~$5, $\Pi~=~$2 bits/s/Hz, $P_{\max, \text{BS}}~=~$25~\text{dBm}, $P_{\text{I}}~=~10~\text{dBm}$, and $M~=~$16. 
\par It is shown that upscaling the number of RIS elements promotes the average system EE for all baselines in Fig. \ref{EE_elements}. In fact, enlarging the number of RIS elements not only improves the system achievable SE (\ref{rate}), but also increases the system power consumption (\ref{p-tot}), on the other hand. For smaller number of RIS elements, the impact of the former case i.e., the system achievable SE dominates the latter one. By incorporating more reflecting elements at the RIS nevertheless, the ascending trend continues yet saturates gradually. This can be justified by pointing out that for larger number of RIS elements incorporated, the domination of the achievable system SE over the consumed system power fades due to huge amount of power consumption at the RIS. Therefore, a practical trade-off can be obtained by considering both the power consumption and achievable SE of the system. Also, as discussed earlier, meta-learning enables capturing the dynamics of AARIS system, mainly originated from the mobility of the users and UAV. Thanks to this advantage, the generalization ability of the learning agent in new environments will be enhanced. To this reason, our \textbf{Proposed MMSAT} algorithm outperforms similar baselines lacking meta-learning. Moreover, the ability of RSMA in better interference management, compared to similar schemes such as NOMA \cite{NOMA1} results in higher average system EE for the \textbf{Proposed MMSAT} algorithm over the \textbf{Baseline 1}. It can be also claimed that because the active RIS can directly amplify and retransmit the incident signals, receivers can better detect and the decode the received signal \cite{multiplicative_fading}. Thus, the \textbf{Proposed MMSAT} algorithm achieves higher system SE (\ref{rate}) and thereby is superior to the \textbf{Baseline 2} with passive RIS, in terms of average system EE.

\par As depicted in Fig. \ref{EE_power}, the more power budget the BS is equipped with, the more average system EE can be attained, yet in a saturating manner. Quantitatively, by progressing the $P_{\text{max,BS}}$ from $5$~dBM to $35$~dBm, about 73\% enhancement yields in average system EE. However, when $P_{max,BS}$ exceeds further, a saturation is observed due to the same reason, we previously discussed. Comparatively, the baselines keep the same order as in Fig. \ref{EE_elements}, owing to the same reasons mentioned there. For instance, the average system EE gained by the \textbf{Proposed MMSAT} algorithm is superior to that related to \textbf{Baseline 3}, which corresponds to a fixed RIS scenario. This observation underscores the impact of mobility in UAV-borne RIS for creating LoS links and extended network coverage. Additionally, by allowing the BS more power budget, more transmit power could be allocated to users, and the system will be more prone to inter-user interference. In this context, a comparison between the \textbf{proposed MSAT} algorithm and \textbf{Baseline 1} signifies the role of interference management and the effectiveness of RSMA over NOMA.
\par By increasing the number of antennas at the BS $N_{\text{BS}}$ as demonstrated in Fig. \ref{N_BS}, the average system EE evidently elevates. This is because of more degree of freedom and spatial multiplexing gain, which enhance the system SE (\ref{rate}). Comparatively, the \textbf{Baseline 2} with active RIS outperforms the \textbf{Baseline 3} with passive mode from the average system EE standpoint. This can be argued that by upscaling the number of BS antennas, although the AARIS system consumes more power, the system SE enhancement dominates and hence, the average system EE improves, as the consequence.




\begin{figure}[ht] 	
	\centering
	\includegraphics[width=0.35\textwidth]{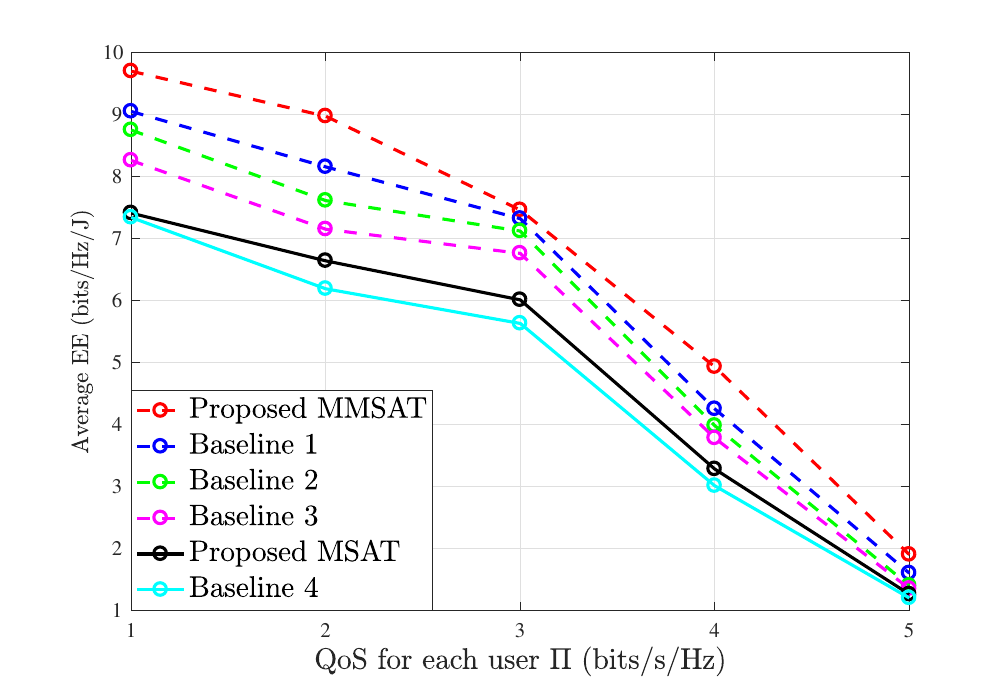}
	\caption{\small Impact of users' QoS threshold on overall average system EE.}
	\label{QoS} 
\end{figure}
By increasing the QoS threshold of users, a decreasing trend for overall system EE is illustrated in Fig. \ref{QoS}. For example, consider that for $\Pi=1$ bits/s/Hz, the \textbf{Proposed MMSAT} algorithm outperforms the \textbf{Baseline 1} for about 8\%, whereas for $\Pi=5$ bits/s/Hz, this superiority degrades to about 2\%. In fact, by guaranteeing more QoS threshold for each user as the constraint C$_2$ indicates in $\mathcal{P}_{0}$, the solution space of the optimization problem $\mathcal{P}_0$ becomes more limited, which therefore degrades the overall system EE as its objective function. Another observation perceived from this figure corresponds to the gap between the baselines, which gets tighter as the user QoS threshold increases.
\vspace*{0.1em}

\begin{table}[t!] 
	\vspace{-1 em}
	\centering
	\caption{Simulation Parameters}
	\label{tab:my_label}
	\vspace*{5pt}
	\begin{tabular}{ |l|l||l|l| }
		\hline
		\textbf{Parameter} & \textbf{Value} & \textbf{Parameter} & \textbf{Value} \\
		\hline
		\hline		
		\hline
		$W_u$ \cite{Rotary_wing_UAV}  & 20 N &
		$\Omega_u$ &  $ 300~\text{rad/s}$\\ 
		\hline $\zeta$ & $1.225~\textrm{kg/m}^3$ &
		$R_u$ & 0.4~m \\ 
		\hline $\rho$ & $ 0.02 $ &
		$d_u$ & 0.3 \\ 
		\hline $\delta$ & $ 0.05 $ &
		$\iota$ & 0.1 \\
		\hline $P_b$ & $79.85~\textrm{Watt}$  &$\nu$& 1.25  \\
		\hline $P_i$ & $88.63~\textrm{Watt}$ &
		$v_{u,i}$ & $4.03~\text{m/s}$ \\
		\hline $P_c$ & $-10~\textrm{dBm}$ &
		$A_u$ & $0.503~\textrm{m}^2$ \\
		\hline $P_\textrm{DC}$ & $-5~\textrm{dBm}$ &
		$\sigma^2_z=\sigma^2_k$ & $-80~\text{dBm}$ \\
		\hline $a_{\textrm{max},u}$ & $6~\text{m/s}^2$ &
		$a_\textrm{max}$ & $20~\textrm{dB}$ \\
		\hline $T$ & $40~s$ &
		$\tau$ & $0.1$ \\
		\hline $P^\textrm{Cir}_k$ & $5~\textrm{mWatt}$ &
		$\Pi_k$ & $2~\text{bits/s/Hz}$ \\
		\hline $P_\textrm{max,BS}$ & $100~\textrm{dBm}$ &
		$C_0$ & $0.001$ \\
		\hline ${K_{\text{BS},u}} = {K_{u,k}}$ & $3~\textrm{dB}$ &
	    $V_{\text{max}, u}$&$10~\text{m/s}$ \\
		\hline $\alpha_{\text{BS},u} = \alpha_{u,k}$& $3$ & $P_\textrm{I}$		 & $10~\textrm{dBm}$ \\
		\hline $\eta$ & $0.8$ & $K$& $3$\\
		\hline $\epsilon$ & $0.8$ & & \\
		\hline
	\end{tabular}
	\label{tab2}
	\vspace{-0.75 em}
\end{table}
\vspace{-1 em}
\section{Conclusions}
This paper investigated the performance of an AARIS system, where the BS employed RSMA for efficient interference management and was assisted by a UAV-borne active RIS. We proposed a real-time and adaptive resource management scheme in this system based on meta reinforcement learning. Through extensive simulations, we found that in spite of more energy consumption, the deployment of an active RIS on the UAV improves the overall system EE. Additionally, a better EE trade-off was characterized for RISs with smaller sizes. 
\bibliographystyle{IEEEtran}
\bibliography{References_AAIRS}
\end{document}